\documentclass[aps,pra,twocolumn]{revtex4}

\usepackage{amsmath}
\usepackage{amssymb}
\usepackage{mathrsfs}
\usepackage{verbatim}
\usepackage{epsfig}
\usepackage{color}
\usepackage[bookmarks]{hyperref}
\usepackage{graphicx}
\usepackage{epstopdf}

\graphicspath{{.}{./EPS/}}

\newcommand* {\bra}[1]{\ensuremath{\langle {#1} |}}
\newcommand* {\ket}[1]{\ensuremath{| {#1} \rangle}}

\newcommand{\hbrho}{\hat{\boldsymbol{\rho}}}

\begin{document}
\title{Hybrid quantum repeater based on resonant qubit-field interactions}

\author{J\'ozsef Zsolt Bern\'ad}
\email{zsolt.bernad@um.edu.mt}
\affiliation{Institut f\"{u}r Angewandte Physik, Technische Universit\"{a}t Darmstadt, D-64289, Germany}
\affiliation{Institute for Solid State Physics and Optics, Wigner Research Centre, Hungarian Academy of Sciences, P.O. Box 49, 
H-1525 Budapest, Hungary}
\affiliation{Department of Physics, University of Malta, Msida MSD 2080, Malta}

\date{\today}

\begin{abstract}

We propose a hybrid quantum repeater based on ancillary coherent field states and material qubits coupled to optical cavities. For this purpose, 
resonant qubit-field interactions and postselective field measurements are determined which are capable of realizing all necessary two-qubit operations
for the actuation of the quantum repeater. We explore both theoretical and experimental possibilities of generating near-maximally-entangled qubit
pairs ($F>0.999$) over long distances. It is shown that our scheme displays moderately low repeater rates, between $5 \times 10^{-4}$ and $23$ 
pairs per second, over distances up to $900$ km, and it relies completely on current technology of cavity quantum electrodynamics.
\end{abstract}

\maketitle

\section{Introduction}

\begin{figure*}[ht!]
 \begin{center}
 \includegraphics[width=.79\textwidth]{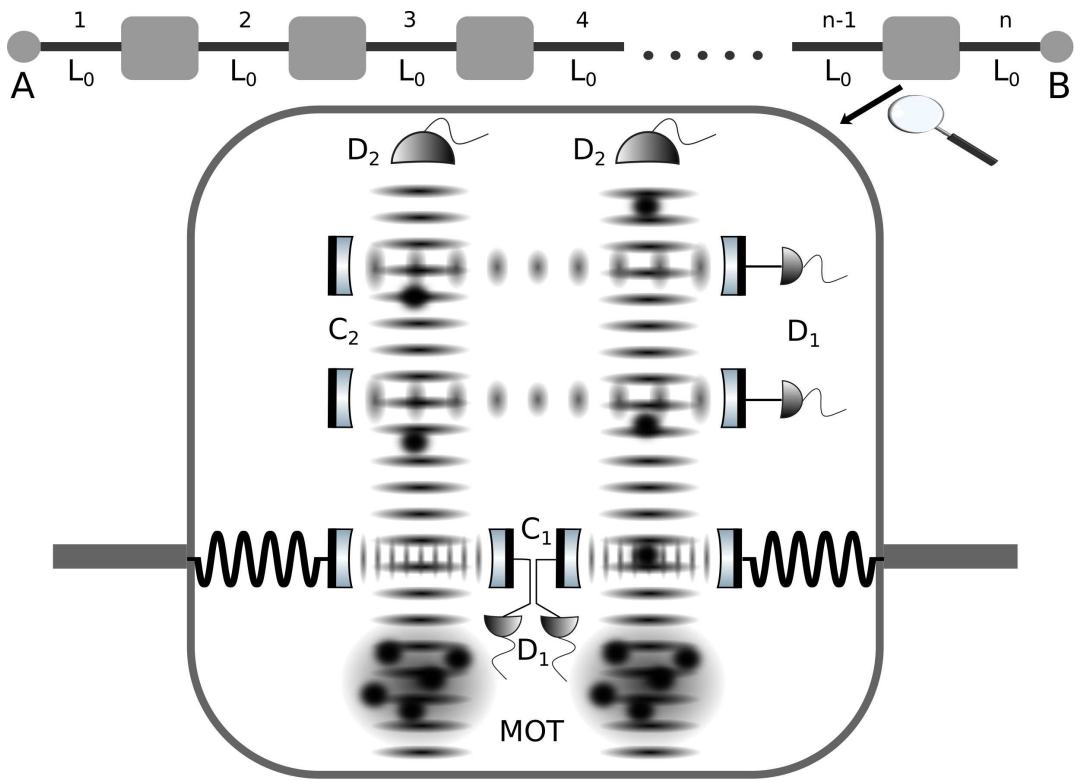}
\caption{Schematic representation of a cavity QED based quantum repeater protocol. The total distance between the end points $A$ and $B$ 
is divided in $n$ elementary links with length $L_0$. At each node there are four cavities: For the entanglement generation protocol 
there are two $C_1$ cavities, in which single qubits can interact with single-mode electromagnetic fields; and 
for the implementation of unambiguous Bell measurements there are two $C_2$ cavities, 
in which two qubits can interact symmetrically and simultaneously with the single-mode electromagnetic fields. The entanglement purification 
protocol is implemented with cavities $C_2$, such that 
there is only one qubit present in the cavity during the qubit-field interaction. The atoms (black dots) implementing 
the qubits are collected from a magneto-optical trap (MOT) and transferred into an optical conveyor belt. The conveyor belt can move the qubits in 
and out of the cavities. There are also two type of detectors: Four $D_1$ detectors realize postselective field measurements for all three 
building blocks of the quantum repeater, and two $D_2$ detectors measure the states of the qubits for both the purification protocol and 
the Bell measurement.}
   \label{setup}
 \end{center}
\end{figure*}

Distribution of well-controlled entanglement over long distances is essential for quantum communication \cite{Nielsen, Kimble}. In practice,
quantum channels connecting spatially separated quantum systems are subject to information loss. For example, direct transmission of photons
via optical fibers, quantum channels, is limited, at best, to a few hundred km \cite{Gisin, Inagaki}. In the case of free space channels, there
are new developments in satellite-Earth-based entanglement distribution \cite{Yin}, which go beyond $1000$ km, though terrestrial free 
space quantum communication has a limitation of a few hundred km due to the curvature of Earth \cite{Ursin}. As straightforward amplification is 
not an option in quantum
communication due to the no-cloning theorem \cite{Dieks, Zurek}, one possibility is to use the quantum repeater 
protocol \cite{Briegel, Dur}, which improves the low success rates. Here, the total distance between the quantum systems is divided into 
smaller distances, i.e., elementary links, with repeater nodes in between. There are already various proposals 
for quantum repeaters and impressive experimental efforts are being made for their implementations \cite{Sangouard}.


An interesting proposal of van Loock {\it et al.} \cite{Loock1,Loock2,Loock3}, a hybrid quantum repeater, uses coherent states to distribute 
entanglement between the nodes of the quantum repeater. This scheme has the advantage that a high repetition rate can be achieved mainly due
to almost unit efficiencies of homodyne photodetection of coherent states, in contrast to low efficiencies of single-photon detectors.
Furthermore, for this type of quantum repeater every logical two-qubit gate is realized with the help of qubit-field interactions within cavity 
quantum electrodynamics (QED) scenarios \cite{Gonta1, Gonta2}. These proposals are based on far-off-resonant qubit-field interactions, which 
impose limitations on the orthogonality of the field states involved in postselective homodyne measurements, thus affecting the fidelity of the 
entangled states. In order to have unit 
fidelities in these approaches, long interaction times or high numbers of mean photons are required. If the interaction times are much longer
than the characteristic times of dipole transitions, then the qubit-field interaction model has to be extend beyond the dipole approximation. 
High numbers of mean photons usually are not an experimental issue; however, to obtain better fidelities one has to increase 
the mean photon number. In this case, the justification of the far-off-resonant model requires significantly increased difference between 
the frequency of the qubit transition and the frequency of the single-mode field, which means that the cavities in use have to have 
adjustable frequencies. It has been demonstrated by us
that this hurdle can be overcome by resonant qubit-field interactions \cite{Bernad1}. As a continuation of this work 
we have proposed building blocks for a hybrid quantum repeater, which is completely based on cavity QED and resonant qubit-field 
interactions \cite{Torres, Ludwig, TorresA}.

In this paper, we go one step further and propose an entire quantum repeater scheme by assembling all three building blocks. 
These building blocks are based on two-level atomic ensembles, single-mode cavities, postselective field measurements, and input coherent states
of the radiation field; hence, they are compatible with each other. Besides assembling a quantum repeater, we also extend our analysis to two 
building blocks. In our previous article 
\cite{Bernad1}, we have studied entanglement generation between two spatially separated material qubits, and now we augment this approach 
with the reflection of photons from the surface of the cavities, a source of decoherence, and we replace 
minimum-error field measurements with balanced homodyne photodetection. We also review the entanglement swapping protocol, 
because in our previously proposed scheme the success probability is found to be less than one 
\cite{TorresA}. In this paper, we present a new set 
of field measurements, which postselect deterministically the four Bell states. The purification 
protocol of Ref. \cite{Ludwig} is applied straightforwardly to output states of the entanglement generation block.

Another aim of this paper is to calculate the average rates of near-maximally-entangled pairs per second between the end points of 
the repeater chain. We focus on a standard quantum repeater scheme \cite{Briegel} and avoid new generation of schemes, for now \cite{Lukin}. 
As a result, the performance of the presented quantum repeater it is expected to be moderate. The rate analysis 
will be done in the context of some current experimental setups with the following assumptions: Qubits do not decay and detectors have unit efficiency.
 
This paper is organized as follows. In Sec. \ref{Prot+Setup}, we present the scheme of the hybrid quantum repeater. In Subsec. \ref{ent}, 
we introduce the theoretical model for the building block of entanglement generation and determine the form of the entangled state generated 
between the repeater nodes. This result
is employed in Subsec. \ref{pur}, where the entanglement purification protocol and its theoretical model are discussed. In Subsec.
\ref{swap}, we present the entanglement swapping protocol and its theoretical model. In Sec. \ref{experiment}, we collect some experimental 
setups and data, which may play a role in the future implementation of the proposal. Based on these experimental setups we determine the 
repeater rates of near-maximally-entangled pairs separated by a distance of $100$ km. 

\section{Protocol}
\label{Prot+Setup}

In this section, we discuss a quantum repeater, which is based on models of cavity quantum electrodynamics. We follow the seminal idea of
Ref. \cite{Briegel}, where entanglement over distance $L$ is created by dividing the distance into $n$ elementary links and inserting 
nodes at their intersection. Thus, each node is connected by elementary links with length $L_0=L/n$ either to neighboring nodes 
or to the end points of repeater chain. The quantum 
repeater presented here consists of three building blocks: Entanglement is generated between neighboring nodes, then with the help of an 
entanglement purification one can purify the effects of any kind of decoherence, and finally entanglement swapping is applied to increase the
distance of shared entanglement. The physical resources of this quantum repeater are atoms, optical conveyor belts, optical or microwave 
cavities with initially prepared coherent states, state-selective detectors for the qubits and postselective field measurements 
implemented by balanced homodyne photodetection. The main
physical mechanism for the realization of two-qubit operations is the resonant qubit-field interaction in dipole and rotating-wave 
approximation. In this context, we employ the Jaynes-Cummings-Paul and Tavis-Cummings models with an interaction time in the region of the 
so-called collapse phenomena \cite{Schleich}. Furthermore, we consider that the initial coherent states have large mean photon numbers 
$\bar n \geqslant 100$.

In Fig. \ref{setup}, we display the sketch of the setup. The status of some current experimental settings, which are strongly related to our proposed
setup, is thoroughly discussed in 
Sec. \ref{experiment}. In the first step, the qubits (implemented by atoms) are loaded from a magneto-optical
trap (MOT) into a dipole trap, which can be set into motion, thus realizing an optical conveyor belt. These qubits interact with single-mode
electromagnetic fields in cavities denoted by $C_1$, which are coupled by optical fibers to neighboring nodes. The emerging fields propagate
to neighboring nodes, where they interact with the local qubits and field measurements ($D_1$ detectors) 
are performed in order to postselect entangled states between qubits separated by an elementary link. 
Afterward, entanglement purification is implemented with the help of qubit-field interactions
in cavities denoted by $C_2$, postselective field measurements ($D_1$ detectors), and qubit measurements ($D_2$ detectors). 
Finally, two cavities (denoted by $C_2$), where simultaneously 
two qubits can interact with the single-mode of the radiation field, and two postselective field measurements are used to generate unambiguous 
and noninvasive Bell measurements. These Bell measurements realize entanglement swapping between the nodes, and after applying them 
in every node we are able to generate near-maximally-entangled pairs between the end points of the repeater chain.  

In the subsequent subsections, we discuss in more detail the quantum electrodynamical models of all three building blocks. The main aim is to 
explore the possibilities of generating near-maximally-entangled pairs with high success probabilities.

\subsection{Entanglement generation between neighboring nodes}
\label{ent}

An elementary link between two neighboring nodes is modeled by two spatially separated cavities $A$, $B$ 
and a long optical fiber connecting them. The qubits in the
nodes are two atoms in conveyor belts with ground states $\ket{0}_i$ and excited states $\ket{1}_i$ ($i \in\{A, B\}$). In the first step, 
the single-mode radiation field of cavity $A$ interacts with qubit $A$. This is followed by the leakage of the single-mode field 
into the optical fiber, the propagation of the radiation field inside the fiber from system $A$ to system $B$, and a leakage of the fiber's 
radiation field into cavity $B$. Finally, the single-mode radiation field in cavity $B$, which is generated by the leakage of the fiber into 
the cavity, interacts with qubit $B$. A postselective measurement on the emerged radiation field in cavity $B$ generates entanglement 
between the two remote qubits. In the subsequent argumentation, we present a minimal model, which is capable of effectively describing this 
physical setup and we analyze its efficiency with respect to the model's parameters in order to generate highly
entangled qubit states with high success probabilities.

We consider a model where both qubits and cavities are similar. There are three main time scales in this system: the qubit-field
interaction time $\tau$, the decay time of the cavities $\tau_c$, and the propagation time $T$ in the fiber. We assume that 
$\tau \ll \tau_c \ll T$, which also encodes our intention that the distance between the two qubits is large. In order to avoid
spontaneous decay of qubit $A$ during the long propagation time, one can coherently transfer the population of the states involved in the 
interaction to radioactively stable electronic levels, which are assumed to not decay during the whole process.
Furthermore, we also consider that the frequency of the  
a single-mode radiation field is in resonance with the qubits' transition frequency. In the dipole and rotating-wave 
approximation the Hamiltonian for the qubit-field interaction is ($\hbar=1$) \cite{Jaynes,Paul}
\begin{eqnarray}
 \hat{H}_1&=& \begin{cases}
         \omega_c \hat{\sigma}^A_z/2+g \hat{a} \hat{\sigma}^A_+ +  g 
\hat{a}^{\dagger} \hat{\sigma}^A_- , \,\,\,\,\,  t \in [0,\tau],\\
         \omega_c \hat{\sigma}^B_z/2+ g \hat{b} \hat{\sigma}^B_+ +  g 
\hat{b}^{\dagger} \hat{\sigma}^B_-, \,\,\,\,\, t \in [T', T'+\tau], 
      \end{cases} \nonumber \\
      T'&=&\tau+T+2\tau_c,\nonumber
\end{eqnarray}
where $\hat{\sigma}_z=\ket{1}_i \bra{1}-\ket{0}_i \bra{0}$, $\hat{\sigma}^i_+=\ket{1}_i \bra{0}$ and $\hat{\sigma}^i_-=\ket{0}_i \bra{1}$ 
($i \in\{A, B\}$). The coupling constant $g$ characterizes the strength of the dipole interaction of the qubits with the single-mode field, 
and thus $2g$ is the vacuum Rabi splitting. $\omega_c$ is the frequency of the single-mode fields in both cavities and 
also the transition frequency of the qubits' state. $\hat{a}$ and $\hat{b}$ ($\hat{a}\dagger$ and $\hat{b}\dagger$ ) are 
the annihilation (creation) operators of the field mode in cavities $A$ and $B$. 

The optical fiber is considered to be a single-mode fiber with frequency $\omega$ and its modes which can couple to the cavities 
form a frequency band $(\omega-\delta \omega,\omega+\delta \omega)$. This usually means that there is only one wave function considered as
the solution of the Helmholtz equation in the cross section of the fiber and many along the length of the fiber \cite{Snyder}. 
In the rotating-wave approximation, i.e., 
$\delta \omega \ll \omega$, the 
interaction Hamiltonian between the single-mode cavities and the
fiber modes is 
\begin{eqnarray}
 \hat{H}_2= \begin{cases} \sum_{i} \kappa_{i,A} 
\hat{a}^\dagger_{i} \hat{a}+ \kappa^*_{i,A} 
\hat{a}^\dagger \hat{a}_i, \,\,\,\,\, t \in [\tau, \tau + \tau_c],\\
         \sum_{i} \kappa_{i,B} 
\hat{a}^\dagger_{i} \hat{b}+ \kappa^*_{i,B} 
\hat{b}^\dagger \hat{a}_i, \,\,\,\,\, t \in [\tau+T, \tau+T+\tau_c], 
      \end{cases} \nonumber
\end{eqnarray}
where $\kappa_{i,A}$($\kappa_{i,B}$) describes the coupling between the single-mode of cavity $A$ ($B$) and the $i$th mode of the fiber.  
$\hat{a}_{i}$ ($\hat{a}^\dagger_{i}$) is the annihilation (creation) operator of the $i$th mode of the fiber. 

Another important phenomena is the photon loss during the propagation of the radiation field through the optical fiber. We consider a model,
where each mode of the fiber is described by a damped harmonic oscillator and the decay rates are equal. The decaying mechanism is given
by the master equation
\begin{equation}
 \frac{d\hat{\rho}}{dt}=-\frac{\gamma}{2} \sum_i \big(\hat{a}^\dagger_i \hat{a}_i \hat{\rho}-2 \hat{a}_i \hat{\rho} \hat{a}^\dagger_i+ \hat{\rho}
\hat{a}^\dagger_i \hat{a}_i\big)=\mathcal{L}\hat{\rho}, \nonumber
\end{equation}
for propagation times $t \in [\tau+\tau_c,\tau+T+\tau_c]$. $\gamma$ is the damping rate, which characterizes the photon loss in
the fiber.  

The free Hamiltonian of the complete radiation field is
\begin{equation}
 \hat{H}_0=\omega_c \left(\hat{a}^\dagger \hat{a}+ \hat{b}^\dagger \hat{b}\right)+ \sum_i \omega_i \hat{a}^\dagger_i \hat{a}_i, \nonumber 
\end{equation}
with $\omega_i \in (\omega-\delta \omega,\omega+\delta \omega)$ being the frequency of the $i$th fiber mode.

The first purpose of this subsection is to investigate the time evolution of the whole setup described by
\begin{equation}
 \frac{d\hat{\rho}}{dt}=-i[\hat{H}_0+\hat{H}_1+\hat{H}_2, \hat{\rho}]+\mathcal{L}\hat{\rho}.
 \label{evolution}
\end{equation}

Our main strategy is to split the above evolution into parts and the output state of one part is considered as
input state for the subsequent one. As we have already stated that $\tau_c \ll T$, it is reasonable
to take the whole time evolution equal to $2\tau+T$ and thus considering the leakages as almost instantaneously occurring effects 
in regard to the propagation time of the radiation field inside the fiber. Therefore, the time evolution can be split into three parts: 
the qubit-field interaction in cavity $A$, the leakages and the photon loss during the propagation, and the qubit-field interaction in cavity $B$.  

First, we investigate the time evolution in cavity $A$ with an initial state 
\begin{equation}
 \ket{\Psi_0}= \ket{0}_A \ket{\alpha}_A,
\label{initialA}
\end{equation}
where the single-mode field is in the coherent state
\begin{equation}
  \label{chst}
 \ket{\alpha}=\sum_{n=0}^\infty 
  e^{-\frac{|\alpha|^2}{2}}
 \frac{\alpha^n}{\sqrt{n!}}
 \ket n,
  \quad\alpha=\sqrt{\bar n}\,e^{i\phi}.
\end{equation}
The other modes (fiber and cavity $B$) of the radiation field are considered to be in the ground state. The initial state of the qubit $B$ is 
not taken into account yet, because it will be prepared after the elapsed time $\tau+T$ and thus qubit $B$ will interact with the 
emerging field in cavity $B$ right after its preparation procedure. Time evolution
of Eq. \eqref{evolution} for times $0 \leqslant t \leqslant \tau $ and initial condition \eqref{initialA} is based on the
solutions of the resonant Jaynes-Cummings-Paul model:
\begin{eqnarray}
  &&\ket{\Psi(t)}=e^{-\frac{|\alpha|^2}{2}} \sum^\infty_{n=0} \Big( \cos(g\sqrt{n}t) \frac{\alpha^n}{\sqrt{n!}} \ket{0}_A e^{i \omega_c t/2} \nonumber \\
  &&-i \sin(g\sqrt{n+1}t) \frac{\alpha^{n+1}}{\sqrt{(n+1)!}} \ket{1}_A e^{-i \omega_c t/2} \Big) e^{-i \omega_c n t} \ket{n}_A. \nonumber
\end{eqnarray}
In the following discussion we focus on large mean photon number $\bar n \gg 1$ and interaction times $\tau$
such that the Rabi frequency $g \sqrt{n}$ can be linearized around $\bar n$. Thus the obtained joint state of qubit and single-mode field
can be approximated by
\begin{eqnarray}
&&\ket{\Psi(\tau)} \approx \nonumber \\
&&\frac{\ket{0}_Ae^{i \omega_c \tau/2}+\ket{1}_Ae^{-i \omega_c \tau/2} e^{i \phi}}{2}e^{-ig \sqrt{\bar n} \tau/2} 
\ket{\alpha(\tau) e^{-i\varphi}}_A \nonumber \\
&&+\frac{\ket{0}_Ae^{i \omega_c \tau/2}-\ket{1}_Ae^{-i \omega_c \tau/2} e^{i \phi}}{2}e^{ig \sqrt{\bar n} \tau/2} 
\ket{\alpha(\tau) e^{i\varphi}}_A, \nonumber \\
&&\varphi=\frac{g \tau}{2 \sqrt{\bar n}}, \quad \alpha(\tau)=\alpha e^{-i \omega_c \tau},  \label{approx}
\end{eqnarray}
provided that the interaction time $\tau$ fulfills the condition
$g\tau /\sqrt{\bar n} \ll 16 \pi $. This is a time scale below the well-known revival phenomena of the population inversion in the 
Jaynes-Cummings-Paul model \cite{Cummings}.

In the next step, qubit $A$ moves out of cavity $A$ and the single-mode radiation field starts its leakage into the optical 
fiber. In order to deal with the dynamics of the second part, leakage out from cavity $A$, propagation in the fiber, and the leakage out from
the fiber into cavity $B$, we recall and make full advantage of the results derived in Ref. \cite{Bernad1}. The initial condition for the leakage
is given by \eqref{approx}, which we rewrite in a more convenient form and we add the ground state of the fiber:
\begin{eqnarray}
\ket{\psi(t=\tau)}&=&\big(\ket{\Psi_1}_A \ket{\alpha^-}_{A}+ \ket{\Psi_2}_A \ket{\alpha^+}_{A}\big)\prod_{i}\ket{0}_i, \nonumber \\
\ket{\Psi_1}_A &=& \frac{\ket{0}_Ae^{i \omega_c \tau/2}+\ket{1}_Ae^{-i \omega_c \tau/2} e^{i \phi}}{2}e^{-ig \sqrt{\bar n} \tau/2}, \nonumber \\
\ket{\Psi_2}_A &=& \frac{\ket{0}_Ae^{i \omega_c \tau/2}-\ket{1}_Ae^{-i \omega_c \tau/2} e^{i \phi}}{2}e^{ig \sqrt{\bar n} \tau/2}, \nonumber \\
\alpha^-&=& \alpha e^{-i \omega_c \tau-i\varphi}, \quad \alpha^+=\alpha e^{-i \omega_c \tau+i\varphi}. \nonumber
\end{eqnarray}
The solution to \eqref{evolution} for times $\tau \leqslant t \leqslant \tau +\tau_c $ is given by
\begin{eqnarray}
\ket{\psi(t)}&=& \ket{\Psi_1}_A \ket{\alpha^{-}(t)}_A \prod_{i} \ket{\alpha^{-}_i(t)}_i \nonumber \\
&+& \ket{\Psi_2}_A \ket{\alpha^{+}(t)}_A \prod_{i} \ket{\alpha^{+}_i(t)}_i \nonumber
\end{eqnarray}
with
\begin{eqnarray}
 \alpha^{\pm}(t)&=&\alpha e^{\pm i \varphi-i\omega_c \tau} e^{-i \omega_c t-\kappa_A t/2}, \nonumber \\
\alpha^{\pm}_i(t)&=&\frac{\alpha e^{\pm i \varphi-i\omega_c \tau} \kappa_{i,A}}{ \omega_i- \omega_c + i  \kappa_A/2}
\left(e^{-i\omega_i t}- e^{-i \omega_c t-\kappa_A t/2} \right), \nonumber
\end{eqnarray}
and $\kappa_A$ being the cavity $A$'s decay constant (see the appendix in Ref. \cite{Bernad1} for a detailed derivation). Provided that
the leakage time $\tau_c$ is sufficiently long, i.e., $\kappa_A \tau_c \gg 1$, and neglecting the small exponential terms, 
we have that the depletion of the cavity mode is perfect. 

In the following step the propagation of the radiation field from cavity $A$
to cavity $B$ is discussed. The initial condition for Eq. \eqref{evolution} with respect to the propagation is
\begin{eqnarray}
\ket{\psi(\tau+\tau_c)}&=& \ket{\Psi_1}_A \prod_{i} \ket{\alpha^{-}_i}_i 
+ \ket{\Psi_2}_A \prod_{i} \ket{\alpha^{+}_i}_i, \nonumber \\
\alpha^{\pm}_i&=& \frac{\alpha e^{\pm i \varphi-i\omega_c \tau-i \omega_i \tau_c} \kappa_{i,A}}{ \omega_i- \omega_c + i  \kappa_A/2}, \label{alpha}
\end{eqnarray}
which also means that we neglect to follow the evolution of the empty cavity $A$. In order to calculate 
$e^{\mathcal{L} t}\Big( \ket{\psi(\tau+\tau_c)} \bra{\psi(\tau+\tau_c)} \Big) $ for times $\tau +\tau_c \leqslant t \leqslant \tau +\tau_c+T $
we recall the results of Sec. III in
Ref. \cite{Bernad1}. Coherent states and coherences between coherent states evolve as
\begin{eqnarray}
&&e^{\mathcal{L}t} \ket{\alpha^{l}_i}_{i}\bra{\alpha^{k}_i}= \nonumber \\
&&e^{-f_i(t)_{l, k}}
\ket{\alpha^{l}_i e^{-\gamma t/2-i \omega_i t}}_{i}\bra{\alpha^{k}_i e^{-\gamma t/2-i\omega_i t}}, \nonumber \\
&& f_i(t)_{l, k}=(1 - e^{-\gamma t})
\Big(\frac{|\alpha^{l}_i|^2+|\alpha^{k}_i|^2}{2}-\alpha^{l}_i (\alpha^{k}_i)^* \Big),\nonumber
\end{eqnarray}
with $l,k \in \{+,-\}$. We observe that $f_i(t)_{+, +}=f_i(t)_{-, -}=0$ for all $i$. For the other two terms, we have
\begin{eqnarray}
 f_i(t)_{+, -}&=& (1 - e^{-\gamma t}) |\alpha^{+}_i|^2 (1-e^{2i \varphi}), \nonumber \\
 f_i(t)_{-, +}&=& (1 - e^{-\gamma t}) |\alpha^{+}_i|^2 (1-e^{-2i \varphi}), \nonumber
\end{eqnarray}
where we used the relation $|\alpha^{+}_i|=|\alpha^{-}_i|$.
We can now conclude that initial to the leakage out from the fiber into cavity $B$ the joint state of fiber modes and qubit $A$ has the following
form:
\begin{eqnarray}
 &&\ket{\Psi_1}_A \bra{\Psi_1} \prod_{i} \ket{\beta^{-}_i}_i \bra{\beta^{-}_i} + 
 \ket{\Psi_2}_A \bra{\Psi_2} \prod_{i} \ket{\beta^{+}_i}_i \bra{\beta^{+}_i} \nonumber \\
 &&+\ket{\Psi_1}_A \bra{\Psi_2} \prod_{i} e^{-f_i(T)_{-, +}}  \ket{\beta^{-}_i}_i \bra{\beta^{+}_i} \label{leakin}  \\
 &&+ \ket{\Psi_2}_A \bra{\Psi_1} \prod_{i} e^{-f^*_i(T)_{-, +}}  \ket{\beta^{+}_i}_i \bra{\beta^{-}_i}, \nonumber
\end{eqnarray}
where
\begin{equation}
 \beta^{\pm}_i=\frac{\alpha e^{\pm i \varphi} \kappa_{i,A}}{ \omega_i- \omega_c + i  \kappa_A/2}
e^{-i\omega_i T-i \omega_c \tau-\gamma T/2} \label{beta}
\end{equation}
and due to the relation $T \gg \tau_c$, we have also considered that $T+\tau_c \approx T$.

Equation \eqref{leakin} can be considered as an initial condition for \eqref{evolution} and we introduce the decay constant $\kappa_B$ for
cavity $B$. We assume that $\kappa_B \tau_c \gg 1$ and as described in Ref. \cite{Bernad1} we have the following conditions for perfect
leakage into cavity $B$: Choose the coupling constants between the fiber and cavity $B$ in such a way that
\begin{eqnarray}
\kappa_{i,B} &=& \kappa^*_{i,A} = \mid \kappa_i\mid e^{-i\varphi_i},\nonumber\\
e^{2i\varphi_i} &=& \frac{\omega_i - \omega_c + i \kappa_A/2}
{\omega_i - \omega_c - i \kappa_A/2}, \nonumber
\end{eqnarray}
which also yields that $\kappa_A=\kappa_B$, and $\omega_i T$ 
is an integer multiple of $2\pi$. If these conditions are not fulfilled, then many photons are reflected from the surface of the mirror, 
which forms cavity $B$ and connects it with the optical fiber. 

In general, these conditions are hard to realize in current experimental setups and therefore we consider a simple yet detailed enough model,
such that it is able to describe effects of photon reflection from the surface of the mirror. We set $\kappa_{i,A}= \kappa_{i,B}$, 
i.e., $\kappa_A=\kappa_B$ and $\omega_i T$ is an integer multiple of $2\pi$ for all $i$. The last condition
can be fulfilled if the relevant modes of the fiber have approximately a frequency spacing of $c/L_0$, where $L_0$ is the length of the fiber and $c$
is the speed of the light in the fiber. Thus, we can introduce (see Eq. A$(19)$ in Ref. \cite{Bernad1}) 
\begin{equation}
\sqrt{\eta}=\sum_i \frac{|\kappa_{i,A}|^2}{\left(\omega_i- \omega_c + i  \kappa_A/2 \right)^2}, \label{eta}
\end{equation}
the transmittance of the mirror and $\eta$ quantifies the fraction of photons which are not reflected back from the surface of cavity $B$.

After $\tau_c$ time, a part of the propagating radiation field is able to leak into cavity $B$ and the reflected field we consider
as a lost information, hence we trace out the state of the fiber after the reflection. By using the relation
\begin{equation}
 \mathrm{Tr}_{\mathrm{fiber}}\{\ket{\{\alpha_i\}}\bra{\{\beta_i\}}\}=e^{-\sum_i \frac{|\alpha_i|^2+|\beta_i|^2-2\alpha_i\beta^*_i}{2}}
 \nonumber
\end{equation}
with $\ket{\{\alpha_i\}}=\ket{\alpha_1}_1 \otimes \ket{\alpha_2}_2 \otimes \dots $ we obtain the 
following joint state of qubit $A$ and the single-mode of field in cavity $B$:
\begin{eqnarray}
 \hat{\rho}&=&\ket{\Psi_1}_A \bra{\Psi_1} \, \ket{\alpha'^{-}}_B \bra{\alpha'^{-}} + 
 \ket{\Psi_2}_A \bra{\Psi_2} \, \ket{\alpha'^{+}}_B \bra{\alpha'^{+}} \nonumber \\
 &+& F(T,\eta,\varphi)\,\ket{\Psi_1}_A \bra{\Psi_2} \, \ket{\alpha'^{-}}_B \bra{\alpha'^{+}} \nonumber \\ 
 &+& F^*(T,\eta,\varphi)\ket{\Psi_2}_A \bra{\Psi_1} \,\ket{\alpha'^{+}}_B \bra{\alpha'^{-}}, \label{AcavB}
\end{eqnarray}
with
\begin{eqnarray}
\alpha'^{\pm} &=& \sqrt{\eta} e^{-\gamma T/2} \alpha e^{-i\omega_c (\tau+ T)\pm i \varphi},  \nonumber \\
F(T,\eta,\varphi)&=&\exp \{-\sum_i f_i(T)_{-,+} + |\beta^R_i|^2 (1-e^{-2i \varphi})\} \nonumber \\
&=&\exp\{-|\alpha|^2(1-e^{-2i \varphi}) \left(1-\eta e^{-\gamma T}\right)\}, \nonumber
\end{eqnarray}
where we have used Eq. \eqref{alpha} for the relation $\sum_i |\alpha^{\pm}_i|^2=|\alpha|^2$ and Eq. \eqref{beta}
to approximate the reflected average photon number $\sum_i |\beta^{R}_i|^2$ as 
$(1-\eta)\sum_i |\beta^{\pm}_i|^2=(1-\eta) |\alpha|^2 e^{-\gamma T}$.

Finally, we consider the last part of the dynamics, where we include the state of qubit $B$, which moves through cavity $B$ 
after the leakage from the fiber is considered to reach its maximum. The initial condition for \eqref{evolution} is
\begin{eqnarray}
\hat{\rho}\otimes \ket{1}_B\bra{1},
\nonumber
\end{eqnarray}
where $\hat{\rho}$ is defined in Eq. \eqref{AcavB}. In order to obtain the solution for the density matrix of the two qubits and 
the single-mode field we separate the initial condition
into four parts. We make use of the resonant Jaynes-Cummings-Paul model for 
the initial condition $\ket{1}_B \ket{\alpha'^{-}}_B$
and obtain in the coherent state approximation
\begin{eqnarray}
&&\frac{\ket{1}_B e^{-i \omega_c \tau/2}+\ket{0}_B e^{i \omega_c \tau/2} e^{-i \phi}}{2}e^{-ig \sqrt{\bar n} \tau/2} 
\ket{\alpha'^{-} e^{-i\varphi}}_B \nonumber \\
&&+\frac{\ket{1}_Be^{-i \omega_c \tau/2}-\ket{0}_Be^{i \omega_c \tau/2} e^{-i \phi}}{2}e^{ig \sqrt{\bar n} \tau/2} 
\ket{\alpha'^{-} e^{i\varphi}}_B, \nonumber
\end{eqnarray}
where we have considered that $\phi+\varphi \approx \phi$. In a similar way, for the initial condition $\ket{1}_B \ket{\alpha'^{+}}_B$ we get
\begin{eqnarray}
&&\frac{\ket{1}_B e^{-i \omega_c \tau/2}+\ket{0}_B e^{i \omega_c \tau/2} e^{-i \phi}}{2}e^{-ig \sqrt{\bar n} \tau/2} 
\ket{\alpha'^{+} e^{-i\varphi}}_B \nonumber \\
&&+\frac{\ket{1}_Be^{-i \omega_c \tau/2}-\ket{0}_Be^{i \omega_c \tau/2} e^{-i \phi}}{2}e^{ig \sqrt{\bar n} \tau/2} 
\ket{\alpha'^{+} e^{i\varphi}}_B. \nonumber 
\end{eqnarray}

In order to present a clear picture of the obtained density matrix of qubits and single-mode field in cavity $B$, we transform
out the phases acquired during the qubit-field interactions and the propagation phase through the optical fiber
\begin{eqnarray}
  \hat{\rho}(2\tau+T)&=&\hat U(2\tau+T)\hat{\rho}'\hat U^\dagger(2\tau+T) ,\label{intpic} \\
 \hat U(2\tau+T)&=&  e^{-i \omega_c \hat{\sigma}^A_z \tau/2-i \omega_c \hat{\sigma}^B_z \tau/2-i \omega_c \hat{b}^\dagger \hat{b} (2 \tau+T)}. 
 \nonumber
\end{eqnarray}
We introduce the Bell states
\begin{eqnarray}
  \ket{\Psi^\pm}&=&\frac{1}{\sqrt2}\left(\ket{0}_A \ket{1}_B \pm \ket{1}_A \ket{0}_B\right), \label{Bell}
\\
  \ket{\Phi^\pm_\phi}&=&\frac{1}{\sqrt2}\left(e^{-i\phi}\ket{0}_A \ket{0}_B\pm e^{i\phi}\ket{1}_A \ket{1}_B\right), \nonumber
\end{eqnarray}
and the unnormalized states
\begin{eqnarray}
 \ket{\Phi_1}&=&\frac{\ket{\Psi^+}+\ket{\Phi^+_\phi}}{2\sqrt{2}} e^{-ig \sqrt{\bar n} \tau} \ket{\alpha_F e^{-2i\varphi}}_B, \nonumber \\ 
 \ket{\Phi_2}&=&\frac{\ket{\Psi^-}-\ket{\Phi^-_\phi}}{2\sqrt{2}}\ket{\alpha_F}_B, \, 
 \ket{\Phi_3}=\frac{\ket{\Psi^-}+\ket{\Phi^-_\phi}}{2\sqrt{2}}\ket{\alpha_F}_B, \nonumber \\  
 \ket{\Phi_4}&=&\frac{\ket{\Psi^+}-\ket{\Phi^+_\phi}}{2\sqrt{2}}e^{ig \sqrt{\bar n} \tau}  \ket{\alpha_F e^{2i\varphi}}_B, \nonumber
\end{eqnarray}
where $\alpha_F=\sqrt{\eta} e^{-\gamma T/2}  \alpha$. The state of 
qubits and single-mode field in the interaction picture defined in \eqref{intpic} takes the form: 
\begin{eqnarray}
&&\hat{\rho}'= \sum^2_{i,j=1} \ket{\Phi_i} \bra{\Phi_j}+\sum^4_{i,j=3} \ket{\Phi_i} \bra{\Phi_j} \nonumber \\
&&+ F(T,\eta,\varphi) 
\Big(\ket{\Phi_1} + \ket{\Phi_2} \Big) \Big(\bra{\Phi_3} + \bra{\Phi_4} \Big) + \mathrm{H.c.} \nonumber
\end{eqnarray}

In the next step, we briefly investigate the possibility of a field measurement which is capable of realizing conditionally an 
entangled two-qubit state. First, we consider the overlaps
\begin{eqnarray}
 &&F_*=|\langle \alpha_F | \alpha_F e^{-2i\varphi} \rangle_B|=|\langle \alpha_F e^{2i\varphi} | \alpha_F \rangle^*_B| \label{ortcond} \\
 =&&|\exp\{-\eta  e^{-\gamma T} |\alpha|^2(1-e^{-2i \varphi})\}| \approx e^{- \eta  e^{-\gamma T} \frac{g^2\tau^2}{2}}, \nonumber 
\end{eqnarray}
where we have used the relation $\varphi=\frac{g \tau}{2 \sqrt{\bar n}}$. The approximation holds for $g\tau \ll \sqrt{\bar n}$ and shows that the overlap
nearly vanishes for interaction times $\tau \geqslant 4\sqrt{e^{\gamma T}/\eta}/g $. In order to ensure that \eqref{ortcond} is almost zero, i.e.,
$\ket{\alpha_F}_B$ is orthogonal to $\ket{\alpha_F e^{2i\varphi}}_B$ and $\ket{\alpha_F e^{-2i\varphi}}_B$, and the 
coherent state approximation is still valid the interaction times have to fulfill the  following condition
\begin{equation}
  4 \sqrt {\frac{e^{\gamma T}}{\eta}} \leqslant g\tau\ll50 \sqrt{\overline n}.
 \label{tau}
\end{equation}
This condition shows clearly the destructive effects of the photon loss during propagation and the photon reflection from the surface
of cavity $B$, i.e., the left-hand side of Eq. \eqref{tau} is the smallest when $\gamma T=0$ and $\eta=1$, which correspond to lossless
propagation or no propagation and perfect leakage into cavity $B$. Thus, for interaction times, 
which fulfill the conditions in \eqref{tau}, there is a postselective field measurement of $\ket{\alpha_F}$ 
by means of balanced homodyne photodetection \cite{Lvovsky} (see also our discussion in Ref. \cite{Torres}), which is able to prepare the 
two-qubit state
\begin{eqnarray}
 \hat{\rho}_{AB}&=&\frac{(1+x)\ket{\Psi^-}\bra{\Psi^-}+(1-x)\ket{\Phi^-_\phi}\bra{\Phi^-_\phi}} {2} \nonumber \\
 &+&\frac{iy\ket{\Phi^-_\phi}\bra{\Psi^-}-
 iy\ket{\Psi^-}\bra{\Phi^-_\phi}}{2}, \label{entgen} \\
 x&=&\exp\{-\bar n \big[1-\cos(2 \varphi) \big] \left(1-\eta e^{-\gamma T}\right)\} \nonumber \\
  &\times& \cos \Big [ \bar n \sin(2 \varphi)\left(1-\eta e^{-\gamma T}\right)  \Big], \nonumber \\
  y&=&\exp\{-\bar n \big[1-\cos(2 \varphi) \big] \left(1-\eta e^{-\gamma T}\right)\} \nonumber \\
  &\times& \sin \Big [ \bar n \sin(2 \varphi)\left(1-\eta e^{-\gamma T}\right)  \Big], \nonumber
\end{eqnarray}
with success probability
\begin{equation}
P_{\mathrm{Gen}}=0.5. \label{entgenprob} 
\end{equation}
 
\begin{figure}[ht!]
 \begin{center}
 \includegraphics[width=.49\textwidth]{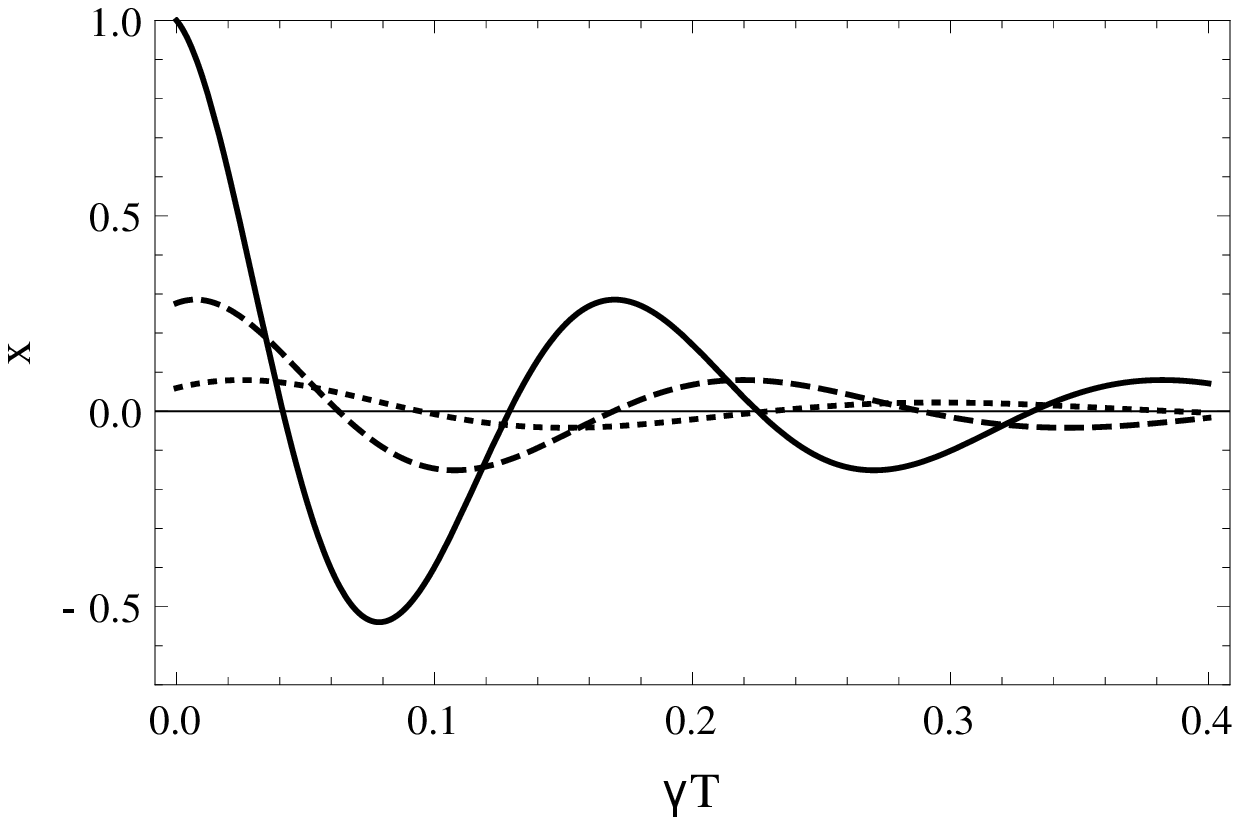}
 \includegraphics[width=.49\textwidth]{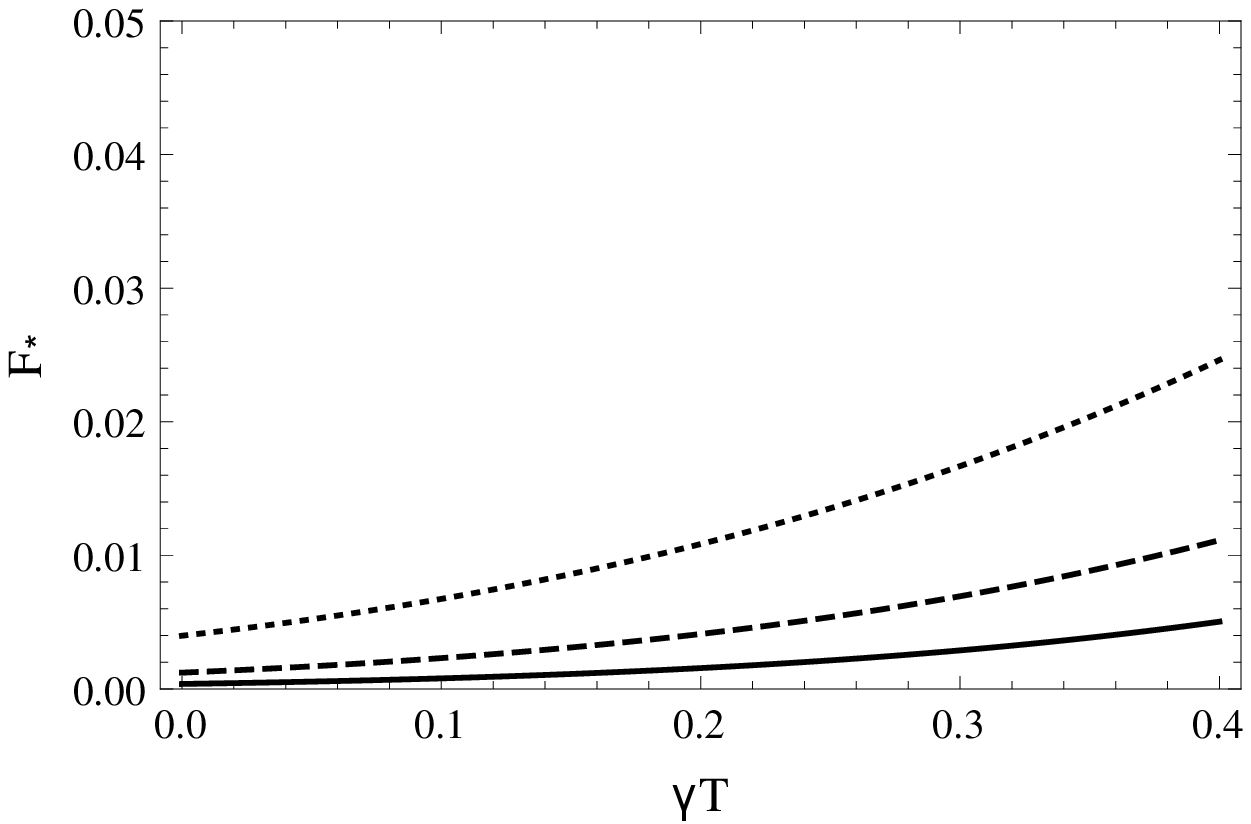}
\caption{
Top panel: Parameter $x$ of the state in Eq. \eqref{entgen} as a function of $\gamma T$. 
Bottom panel: The overlap of the field states, $F_*$ in Eq. \eqref{ortcond}, as a function of $\gamma T$. 
The interaction time was set to $\tau=4/g$ with mean photon number $\overline n =100$. Three curves are presented for 
different values of $\eta$: $1.0$ (full line), $0.85$ (dashed line), and $0.7$ (dotted line).}
   \label{fig1}
 \end{center}
\end{figure}

The introduced parameter $x$ is going 
to be the only crucial ingredient in the analysis of the repeater rates, which we are going to show during the discussion of the entanglement purification procedure.
It is worth to note that the concurrence of the state in Eq. \eqref{entgen} is $\sqrt{x^2+y^2}$ \cite{Wootters}. 

The last part of this section is devoted to the
analysis of $x$ as a function of $\eta$ and $\gamma T$. Figure \ref{fig1} shows the limitations of entanglement generation between the two remote
qubits. We require in general that the absolute value of $x$ in Eq. \eqref{entgen} to be close to unity and on the other hand that the 
overlap between components of the field state are zero, which is a necessary condition for successful postselection of \eqref{entgen}. These two 
requirements are hard to hold, for cases when the photon loss in the fiber is high and there is a big reflection rate from the surface of cavity 
$B$. The separation angle $\varphi$ between the coherent state during the qubit-field interactions can not be too large ($\approx \pi/4$), 
because then $|x|$ is too small, but when $\varphi$ is too small then the overlap between the field states is not vanishing. This is an
optimization problem with competing objectives and the ideal $\varphi$, which is determined by the interaction time (see Eq. \eqref{approx}), has to be found for every
experimental scenario independently. The concurrence of the generated state as an indicator of the entanglement quality is nothing else than the upper envelope of the curves in Figure \ref{fig1}.

We conclude at the end of this subsection that entanglement generation between two nodes connected by a optical fiber is obtained. The resulting
two-qubit states are not perfect Bell states, but they are entangled unless the parameter $x$ is equal to zero. In the following subsection, 
we investigate these states as input states for the entanglement purification protocol. 

\subsection{Entanglement purification}
\label{pur}

In this subsection we present an entanglement purification protocol, which is capable of increasing the degree of entanglement
of the state obtained in Sec. \ref{ent} (see Eq. \ref{entgen}). The protocol we have in mind is a recurrence protocol \cite{Bennett,Deutsch},
which works in a recursive way; i.e., it uses two copies of the same state for the next purification step. Thus, it is assumed
that the entanglement generation procedure have resulted in a presumably large ensemble of similarly entangled states between the repeater nodes. 
In our previous work \cite{Ludwig}, we have discussed an implementation and here we briefly recapitulate it.

We consider two qubits $A_1$ and $A_2$ in one node with ground states $\ket{0}_i$ and excited states 
$\ket{1}_i$ ($i \in\{A_1,A_2\}$). These qubits move sequentially through a cavity and interact resonantly with
single-mode field prepared initially in a coherent state $\ket{\alpha}$ (see Eq. \eqref{chst}). We take a general initial state with no 
correlations between the field and the qubits
\begin{equation}
\ket{\Psi_0}= \left (c_{00} \ket{00} + c_{01} \ket{01} + c_{10} \ket{10} + c_{11} \ket{11} \right) \ket{\alpha},
\nonumber
\end{equation}
with the basis $\ket{ij}=\ket{i}_{A_1}\ket {j}_{A_2}$ ($i,j \in \{0,1\}$). The Hamiltonian in the dipole and rotating-wave
approximation reads
\begin{eqnarray}
 \hat{H}&=& \begin{cases}
         \omega_c \hat{\sigma}^{A_1}_z/2+\omega_c \hat{a}^\dagger \hat{a} +g \hat{a} \hat{\sigma}^{A_1}_+ +  g 
\hat{a}^{\dagger} \hat{\sigma}^{A_1}_- , \,\,\,\,\,  t \in [0,\tau],\\
         \omega_c \hat{\sigma}^{A_2}_z/2+\omega_c \hat{a}^\dagger \hat{a} +g \hat{a} \hat{\sigma}^{A_2}_+ +  g 
\hat{a}^{\dagger} \hat{\sigma}^{A_2}_- , \,\,\,\,\,  t \in [\tau, 2\tau], 
      \end{cases} \nonumber 
\end{eqnarray}
where  $\sigma^{i}_z=\ket{1}_i \bra{1}-\ket{0}_i \bra{0}$, $\hat{\sigma}^i_+=\ket{1}_i \bra{0}$ and $\hat{\sigma}^i_-=\ket{0}_i \bra{1}$ 
($i \in\{A_1, A_2\}$). $2g$ is the vacuum Rabi splitting and $\omega_c$ is the frequency of the single-mode field in the cavity and 
also the transition frequency of the qubits' state. $\hat{a}$ ($\hat{a}\dagger$) is the annihilation (creation) operator of the field 
mode in the cavity.

By solving the Jaynes-Cummings-Paul model in sequence for interaction times
characterizing the collapse phenomena \cite{Cummings} and projecting
onto the field state contribution $\ket{\alpha}$ in the full solution by means of balanced homodyne photodetection \cite{Ludwig}, 
one is able to generate the probabilistic two-qubit quantum operation
at each node:
\begin{equation}
  \hat{M}_{A_1,A_2}=\ket{\Psi^-}\bra{\Psi^-}+\ket{\Phi^-_\phi}\bra{\Phi^-_\phi},
 \label{Qgate}
\end{equation}
where these Bell states are defined by Eq. \eqref{Bell} with the indices $A_1=A$ and $A_2=B$. This probabilistic quantum operation 
takes over the role of the controlled-NOT gate employed
in the seminal protocols of Refs. \cite{Bennett, Deutsch}. 

We have demonstrated in Ref. \cite{Ludwig} that for large mean photon number $\bar n=500$, cavity damping, and spontaneous emission of
the qubits (parameter values based on Ref. \cite{Haroche}), the following protocol
is very robust: \\
(I) The quantum operation $M$ is applied locally at each node $A$ and $B$ to the initial state
\begin{equation}
\hbrho=\hat{\rho}_{A_1,B_1} \otimes \hat\rho_{A_2,B_2},
\nonumber
\end{equation}
where both $\hat{\rho}_{A_1,B_1}$ and $\hat\rho_{A_2,B_2}$ have the form of Eq. \eqref{entgen}. After successful
applications of the quantum operations at each node $A$ and $B$, we get the following four-qubit state  
\begin{equation}
\hbrho^{(1)}=
\frac{ \hat{\bf{M}} \hbrho \hat{\bf{M}}^\dagger }{\mathrm{Tr} 
\left\{\hat{\bf{M}}^\dagger \hat{\bf{M}} \hbrho\right\}}
,\quad  \hat{\bf{M}}=\hat{M}_{A_1,A_2}  \hat{M}_{B_1,B_2}.
\nonumber
\end{equation}
(II)  One of the pairs is measured, where the choice of the measured pair is unimportant. 
There are four possible states in which one can find, for example, the pair $(A_2,B_2)$. 
The measurement of one of the states $\ket{ij}_{A_2,B_2}$ with $i,j \in \{0,1\}$ results in
the two-qubit state
\begin{equation}
  \hat{\rho}^{i,j}_{A_1,B_1}=\mathrm{Tr}_{A2,B2}\left\{\ket{ij}_{A_2,B_2} \bra{ij}\hbrho^{(1)}\right\}. 
\end{equation}
(III) In the next step, we apply the unitary operator 
$\hat{U}^i_{A_1} \hat{U}^{j+1}_{B_1}$ at each node to the state $ \rho^{i,j}_{A_1,B_1}$, where
\begin{equation}
\hat{U}^i=
\big(\ket{1}\bra{1}+i\ket{0}\bra{0}\big)\big(\ket{1}\bra{0}+\ket{0}\bra{1}\big)^j. \nonumber
\end{equation}
The final two-qubit state is obtained after a measurement dependent ($A_2$ and $B_2$ qubits found in the states $\ket{ij}_{A_2,B_2}$) unitary transformation
\begin{equation}
  \hat{\rho}^{(1)}_{A_1,B_1}=
  \left(\hat{U}^i_{A_1} \hat{U}^{j+1}_{B_1}\right)
  \hat{\rho}^{i,j}_{A_1,B_1}
  \left(\hat{U}^i_{A_1} \hat{U}^{j+1}_{B_1}\right)^\dagger. \nonumber
\end{equation}
\begin{figure}[t!]
 \begin{center}
 \includegraphics[width=.49\textwidth]{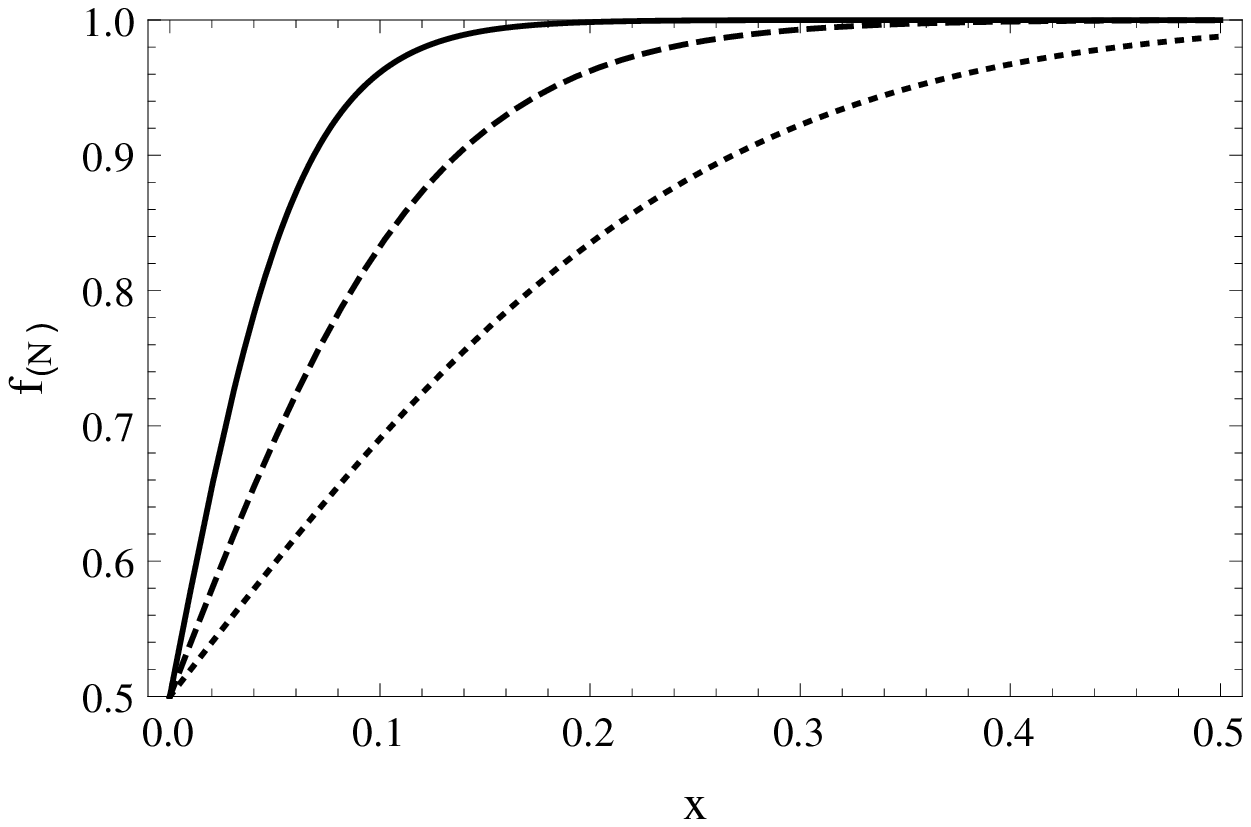}
 \includegraphics[width=.49\textwidth]{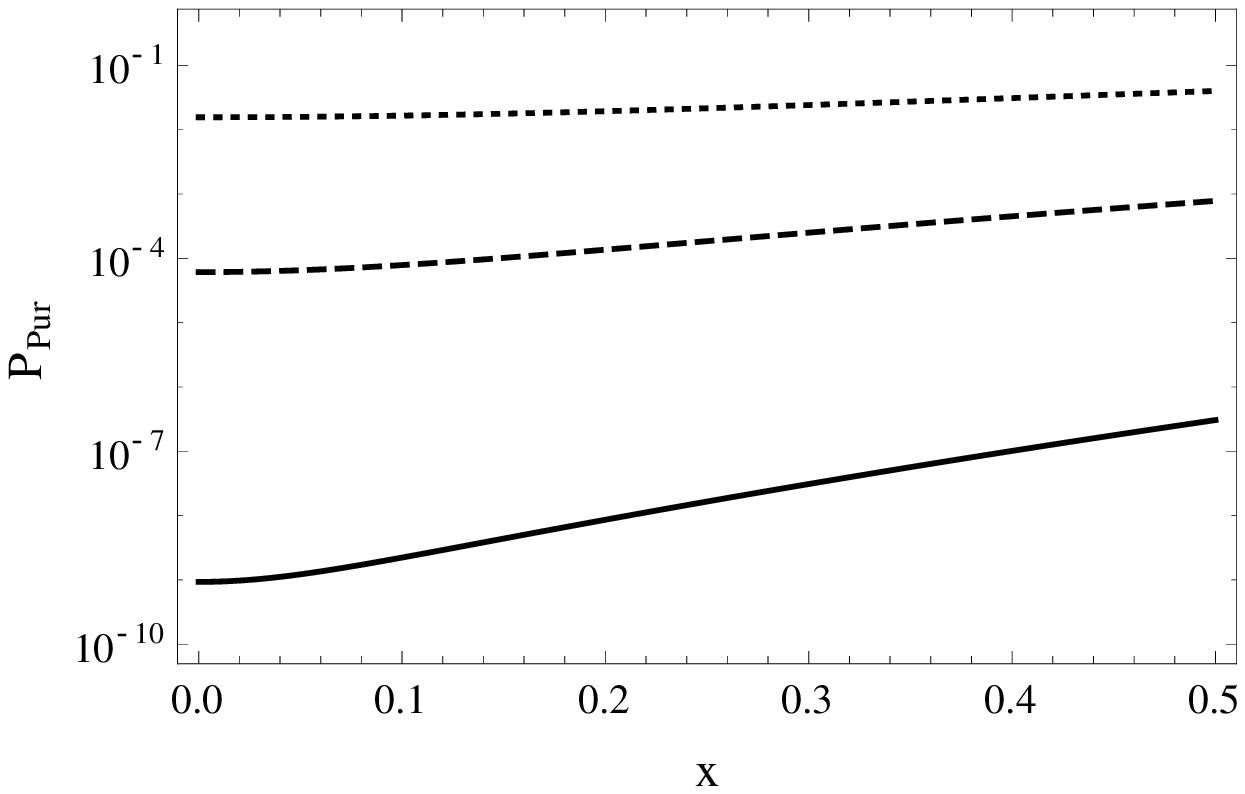}
\caption{
Top panel: The coefficient $f_{(N)}$ in Eq. \eqref{purifyN} as a function of $x$
after $N$ iterations of the purification protocol. 
Bottom panel: Semilogarithmic plot of the overall success probability $P_{\mathrm{Pur}}$ (see Eq. \eqref{purprob}) as a function of $x$.
The value of $x$, a parameter of the state in Eq. \eqref{entgen}, is considered to be not larger than
$0.5$ due to the findings in Fig. \ref{fig1}. Three curves are presented for 
different values of iterations $N$: $4$ (full line), $3$ (dashed line), and $2$ (dotted line).}
   \label{fig2}
 \end{center}
\end{figure}

Now, we recall the result of Sec. \ref{ent} on the state generated between two spatially separated qubits. For simplicity, 
we set the phase $\phi$ in \eqref{entgen} equal to zero and we apply a local unitary transformation at both nodes $A$ and $B$ 
such that $\ket{\Psi^-}$ picks up a global phase and $\ket{\Phi^-} \to \ket{\Phi^+}$. Substituting this state into 
the purification protocol, we get
\begin{eqnarray}
\hat{\rho}^{(1)}_{A_1,B_1}&=&\frac{(1+x)^2}{2+2x^2} \ket{\Psi^-}\bra{\Psi^-} +  \frac{(1-x)^2}{2+2x^2} \ket{\Psi^+}\bra{\Psi^+} \nonumber \\
&+&\frac{y^2}{2+2x^2} \ket{\Psi^-}\bra{\Psi^+}+\frac{y^2}{2+2x^2} \ket{\Psi^+}\bra{\Psi^-}, \nonumber
\end{eqnarray}
with a success probability of $(1+x^2)/4$. After repeating the protocol for $N$ times, i.e., $2^N$ qubit pairs 
were used to get a single two-qubit state, we obtain
\begin{eqnarray}
\hat{\rho}^{(N)}&=&f_{(N)}(x)\ket{\Psi^-}\bra{\Psi^-} +  g_{(N)}(x) \ket{\Psi^+}\bra{\Psi^+} \nonumber \\
&+&h_{(N)}(x)\big( \ket{\Psi^-}\bra{\Psi^+}+\ket{\Psi^+}\bra{\Psi^-} \big), \label{purifyN}
\end{eqnarray}
with success probability
\begin{eqnarray}
 P_{\mathrm{Pur}}&=&\big[P_{(0)} \big]^{2^{N-1}} \big[P_{(1)}\big]^{2^{N-2}} \dots \big[P_{(N-2)}\big]^2 P_{(N-1)},  \nonumber \\
 P_{(k)}&=&\frac{f^2_{(k)}(x)+ g^2_{(k)}(x)}{2}, \label{purprob}
\end{eqnarray}
where
\begin{eqnarray}
 f_{(k+1)}(x)&=&\frac{f^2_{(k)}(x)}{f^2_{(k)}(x)+g^2_{(k)}(x)}, \quad  f_{(0)}(x)=\frac{1+x}{2}, \nonumber \\
 g_{(k+1)}(x)&=&\frac{g^2_{(k)}(x)}{f^2_{(k)}(x)+g^2_{(k)}(x)}, \quad  g_{(0)}(x)=\frac{1-x}{2}, \nonumber \\
 h_{(k+1)}(x)&=&\frac{h^2_{(k)}(x)}{f^2_{(k)}(x)+g^2_{(k)}(x)}, \quad  h_{(1)}(x)=\frac{y^2}{2+2x^2}. \nonumber
\end{eqnarray}
In general, $h_{(N)}(x)$ tends to zero as $N$ increasing and stays constant at the value $0.5$ only when $y=1$. However, when
$y=1$ then $x=0$, which means that the state in  Eq. \eqref{entgen} is not purifiable. Equation \eqref{purprob} shows the overall success probability of $N$ purification rounds,
provided that in the first iteration $2^{N-1}$ qubits, in the second iteration $2^{N-2}$ qubits, and so on, are successfully purified. 

It is worth noticing that we purify in the case of $x>0$ towards the Bell state $\ket{\Psi^-}$ and for 
$x<0$ towards the other Bell state $\ket{\Psi^+}$. In Fig. \ref{fig2}, we show the tradeoff between the entanglement
of the state and the overall success probability after several iterations of the purification protocol. The figures demonstrate that 
the purification protocol is successful only at the expense of the qubit resources due to the low probabilities involved \cite{commentus}. 
For example, $N=4$ iterations with $16$ qubit pairs can purify a wide range of badly entangled states towards a Bell state; 
however, the overall success probability is approximately $10^{-8}$ to $10^{-9}$. For lower number of iterations
we can cover a smaller range of badly entangled states, but with higher overall success probability and better resource management.
We presented a figure only for the coefficient $f_{(N)}$, because $g_{(N)}$ is nothing else than the reflection of 
$f_{(N)}$ about the vertical axis. The absolute value of $x$ is limited
by a value of $0.5$ due to our findings in the Sec. \ref{ent}.

\subsection{Entanglement swapping}
\label{swap}

In an entanglement swapping the goal is to increases the distance of the shared entanglement. 
In other words take three repeater nodes, labeled by the letters $A$, $B$, and $C$, where $A$-$B$ and $B$-$C$ are
neighboring nodes which share a Bell pair. Thus, $B$ has two qubits and by performing a projective Bell measurement on these two qubits
and communicating the results to $A$ and $C$, one can generate a Bell pair between
$A$ and $C$. Applying the swapping protocol to all intermediate nodes results in a Bell pair
between the end points of the repeater chain.

The most important ingredient here is the realization of Bell measurements. We have already introduced and discussed
such a scenario, where we have analyzed the realization of a noninvasive unambiguous Bell measurements \cite{Torres, TorresA}. A noninvasive 
Bell measurement means that the qubits are projected on a Bell state without destroying them. Therefore, measuring later these two-qubit states
one can demonstrate that they are indeed in a Bell state. Our proposal is compatible with the other two protocols presented in 
Secs. \ref{ent} and \ref{pur}, because it is based on the Tavis-Cummings
model \cite{Tavis}, i.e., two material qubits interact simultaneously with the single-mode radiation field inside a cavity, 
and postselective field measurements. However, the 
proposed scheme in Ref. \cite{TorresA} requires special conditions on the mean photon number of the single-mode fields. In 
the subsequent discussion we briefly present the scheme proposed in Refs. \cite{Torres, TorresA} with a different postselective field measurement 
scheme, based also on balanced homodyne photodetection.

Let us consider that two qubits $A$ and $B$ simultaneously move through a cavity and resonantly interact with the single-mode radiation field, 
where the path of the qubits is designed such a way that the dipole couplings $g$ are equal. Thus, the Hamiltonian in the dipole and 
rotating-wave approximation reads
\begin{eqnarray}
 \hat{H}&=&\omega_c \hat{a}^\dagger \hat{a} 
 +g \hat{a} \left( \hat{\sigma}^{A}_+ +  \hat{\sigma}^{B}_+ \right) +  g \hat{a}^{\dagger} \left( \hat{\sigma}^{A}_- + \hat{\sigma}^{B}_- \right)
 \nonumber \\
 &+&\omega_c \left(\hat{\sigma}^{A}_z/2 + \hat{\sigma}^{B}_z/2 \right), \nonumber 
\end{eqnarray}
where  $\sigma^{i}_z=\ket{1}_i \bra{1}-\ket{0}_i \bra{0}$, $\hat{\sigma}^i_+=\ket{1}_i \bra{0}$ and $\hat{\sigma}^i_-=\ket{0}_i \bra{1}$ 
($i \in\{A, B\}$). $\omega_c$ is the frequency of the single-mode field in the cavity and 
also the transition frequency for qubits $A$ and $B$. $\hat{a}$ ($\hat{a}\dagger$) is the annihilation (creation) operator of the single-mode field.

The field is prepared initially in a coherent
state and after the interaction it is postselected by balanced homodyne photodetection. In the next step, the two qubits move through the second cavity and interact
with the single-mode radiation field, prepared also in a coherent state. The emerged state of the field is again postselected. If each of the two 
postselections has two outputs, then there are four possible two-qubit states which are generated in the protocol. The main task is to find
those conditions which allow that these four postselected qubit states are the Bell states.

The initial condition before the first interaction is
\begin{equation}
  \ket{\Psi_0}=
  \Big(a_-\ket{\Psi^-}+
  a_+\ket{\Psi^+}+b_-\ket{\Phi^-}+
  b_+\ket{\Phi^+}
  \Big)
  \ket{\alpha}, \label{initialBell}
\end{equation}
where $\alpha=\sqrt{\bar n}$,
and the Bell states are defined in Eq. \eqref{Bell} with the following adopted notation:
\begin{equation}
 \ket{\Phi^\pm}=\ket{\Phi^\pm_{\phi=0}}. \nonumber
\end{equation}

By solving the resonant Tavis-Cummings
model for interaction times characterizing the collapse phenomena \cite{Cummings} and approximating the field contributions 
one order beyond the coherent
state approximation one obtains (see the appendix in Ref. \cite{TorresA})
\begin{eqnarray}
  \ket{\Psi(\tau)}&\approx &
 \left(a_-\ket{\Psi^-}+b_-\ket{\Phi^-}\right) \ket{\alpha}\nonumber\\
  &+&\frac{a_+-b_+}{2} \left(\ket{\Psi^+}-\ket{\Phi^+_{2\pi\tau}}\right)
  \ket{\alpha_+}\nonumber\\
  &+&\frac{a_+ +b_+}{2}\left(\ket{\Psi^+}+\ket{\Phi^+_{-2\pi\tau}}  \right)
  \ket{\alpha_-}, \label{PsiBell}
\end{eqnarray}
where $\tau$ is a dimensionless parameter of the interaction time equal to $\frac{gt}{\pi \sqrt{4 \bar n +2}}$ and
we have introduced the field states 
\begin{equation}
  \ket{\alpha_\pm}=
  \sum_{n=0}^\infty
  \frac{\alpha^ne^{-\frac{|\alpha|^2}{2}}}{\sqrt{n!}}
  e^{\pm i 2\pi\tau\left[\bar n+1+n-\frac{(n-\bar n)^2}{4\bar n +2} \right]}\ket {n}.
\end{equation}
The collapse phenomena occurs when $1/4 \leqslant \tau \leqslant 3/4$. If $\tau=1/2$, then the field states $\ket{\alpha_\pm}$
have made half a rotation in phase space and lie on the opposite site to the initial coherent state $\ket{\alpha}$, 
i.e., $\langle \alpha | \alpha_\pm \rangle \approx 0$. This is a special
case because the qubit states has the following relation:
\begin{equation}
 \ket{\Phi^+_{\pi}}=\ket{\Phi^+_{-\pi}}=-\ket{\Phi^+}.
\end{equation}
Furthermore, $\ket{\alpha_-}$ ($\ket{\alpha_+}$) rotates clockwise (counterclockwise) during the interaction time on the 
circle with radius $\sqrt{\bar n}$ and
at $\tau=1/2$ $\langle \alpha_- | \alpha_+ \rangle \neq 0$.   

In the next step a postselective measurement on the field state $\hat{\rho}=\mathrm{Tr}\{\ket{\Psi(\tau)} \bra{\Psi(\tau)}\}$ is performed 
with the help of balanced homodyne photodetection. Here, we briefly recapitulate the basics of this measurement, because the arguments
presented in this subsection differ from our former study in Ref. \cite{TorresA}. So, the quantum state of the field, which we want to
measure, interferes with an intense coherent state $\ket{|\alpha_{L}|e^{i\phi_L}}$ of a local oscillator on a $50\%:50\%$ beam splitter. 
The two modes emerging from the beam splitter are directed to two photodetectors which generate an electric current proportional to the 
photon number. The two photocurrents are subtracted and thus by
the difference of photon numbers $n_-$ is measured. Provided that the local oscillator state is intense, i.e., $\mid \alpha_L\mid \gg 1$,  and 
the photodetectors have unit efficiency, the
measurement is equivalent to a projective von Neumann measurement \cite{Lvovsky}. If $a$ and $a^\dagger$ are the annihilation and creation 
operator of the mode to be measured, 
then a quadrature state $\ket{x_{\phi_L}}$ is defined by the relation
\begin{equation}
  \frac{1}{\sqrt{2}} \left(\hat{a}e^{-i\phi_L} + \hat{a}^{\dagger}e^{i\phi_L}\right)
\ket{x_{\phi_L}} = x_{\phi_L}\ket{x_{\phi_L}}.
\end{equation}
A balanced homodyne measurement projects onto the quadrature eigenstate $\ket{x_{\phi_L}}$ according to
the probability distribution
\begin{equation}
P_{\phi_L}\left(x_{\phi_L}\right) =
{\rm Tr}
\{
\hat{\rho} \ket{x_{\phi_L}} \bra{x_{\phi_L}}
\},
\end{equation}
where $x_{\phi_L}=n_-/\sqrt{2|\alpha_L|^2}$. For the case $\tau=1/2$, we consider the following projective measurement,
\begin{equation}
 \hat{\mathcal P}_1=\int^{\infty}_0 \ket{x_{\pi/2}} \bra{x_{\pi/2}} dx_{\pi/2},
\end{equation}
with properties
\begin{eqnarray}
 \hat{\mathcal P}_1 \ket{\alpha}=\ket{\alpha}, \quad \hat{\mathcal P}_1 \ket{\alpha_\pm}=0.
\end{eqnarray}
The projector $ \hat{\mathcal P}_1$ represents a postselective balanced homodyne measurement where the right side of the 
phase space is measured only. 

In the case when the detectors are signaling, we postselect from the joint state of field and qubits \eqref{PsiBell} the state
\begin{equation}
 a_-\ket{\Psi^-}+b_-\ket{\Phi^-}=a_-\ket{\Psi^-}-b_-\ket{\Phi^+_{\pi/2}}.
\end{equation}
Hence, the initial state before the second interaction reads
\begin{equation}
  \ket{\Psi_0}=
  \Big(a_-\ket{\Psi^-}-b_-\ket{\Phi^+_{\pi/2}}
  \Big)
  \ket{\alpha e^{i\pi/2}}.
\end{equation}
After $\tau=1/2$ interaction time, we get the state
\begin{eqnarray}
  &&a_-\ket{\Psi^-} \ket{\alpha'}+b_-\ket{\Phi^+_{-\pi/2}} \Big(\ket{\alpha'_+} + \ket{\alpha'_-}\Big) \nonumber \\
  &&+b_-\ket{\Psi^+} \Big(\ket{\alpha'_-} - \ket{\alpha'_+}\Big),
\end{eqnarray}
where $\alpha'=\alpha e^{i\pi/2}$ and we have used the relation $\ket{\Phi^+_{-\pi/2}}=\ket{\Phi^+_{3\pi/2}}$. The field of the second cavity is
measured with the help of the following projector,
\begin{equation}
 \hat{\mathcal P}_2=\int^{\infty}_0 \ket{x_{0}} \bra{x_{0}} dx_{0},
\end{equation}
with properties
\begin{eqnarray}
 \hat{\mathcal P}_2 \ket{\alpha'}=\ket{\alpha'}, \quad \hat{\mathcal P}_2 \ket{\alpha'_\pm}=0. \label{cav2proj1} 
\end{eqnarray}
The projector $ \hat{\mathcal P}_2$ represents a postselective balanced homodyne measurement where the upper side of the 
phase space is measured only. If the detector signals, then we postselect the two qubit state $\ket{\Psi^-}$ with success probability
$|a_-|^2$ (compare with the initial condition in \eqref{initialBell}). Otherwise, we have
\begin{eqnarray}
 \left(\hat{\mathcal{I}}-\hat{\mathcal P}_2 \right) \ket{\alpha'}&=&0, \nonumber \\
 \left(\hat{\mathcal{I}}-\hat{\mathcal P}_2 \right) \Big(\ket{\alpha'_-} - \ket{\alpha'_+}\Big)&=&0, \label{cav2proj2} \\
 \left(\hat{\mathcal{I}}-\hat{\mathcal P}_2 \right) \Big(\ket{\alpha'_+} + \ket{\alpha'_-}\Big)&=&\Big(\ket{\alpha'_+} + \ket{\alpha'_-}\Big),
 \nonumber
\end{eqnarray}
where $\hat{\mathcal{I}}$ is the identity operator. Thus, in the case of no signaling we postselect 
the two-qubit state $\ket{\Phi^+_{-\pi/2}}$ with success probability $|b_-|^2$.

Let us turn back to the case when the detectors which postselect the field of the first cavity do not signal. In this case, we obtain
the following two-qubit state:
\begin{equation}
 a_+\ket{\Psi^+}+b_+\ket{\Phi^+_{\pi}}=a_+\ket{\Psi^+}-b_+\ket{\Phi^-_{\pi/2}}.
\end{equation}
Thus, before the second interaction, the initial state reads
\begin{equation}
  \ket{\Psi_0}=
  \Big(a_+\ket{\Psi^+}-b_+\ket{\Phi^-_{\pi/2}}
  \Big)
  \ket{\alpha e^{i\pi/2}},
\end{equation}
and after $\tau=1/2$ interaction time, we have the following joint state of field and qubits:
\begin{eqnarray}
  &&-b_+\ket{\Phi^-_{\pi/2}} \ket{\alpha'}+a_+\ket{\Phi^+_{-\pi/2}} \Big(\ket{\alpha'_-} - \ket{\alpha'_+}\Big) \nonumber \\
  &&+a_+\ket{\Psi^+} \Big(\ket{\alpha'_+} + \ket{\alpha'_+}\Big).
\end{eqnarray}
In case of successful signaling when $\hat{\mathcal P}_2$ is applied, the protocol postselects the state
$\ket{\Phi^-_{\pi/2}}$ with success probability $|b_+|^2$; otherwise, i.e., $\hat{\mathcal{I}}-\hat{\mathcal P}_2$ is applied,
we get $\ket{\Psi^+}$ with $|a_+|^2$ success probability.

Let us apply these results to the three-node scheme ($A$, $B$, and $C$ ) explained at the beginning of this subsection. 
The initial state we consider is 
\begin{eqnarray}
\ket{\Psi_0}&=&\ket{\Psi^-}_{AB_1} \otimes \ket{\Psi^-}_{ B_2C}
\\
&=&
-\frac{1}{2} \ket{\Psi^-}_{AC}\ket{\Psi^-}_{B_1B_2}
-\frac{1}{2} \ket{\Phi^+}_{AC}\ket{\Phi^+}_{B_1B_2}
\nonumber\\
&+&\frac{1}{2}\ket{\Psi^+}_{AC}\ket{\Psi^+}_{B_1B_2}
+\frac{1}{2} \ket{\Phi^-}_{AC}\ket{\Phi^-}_{B_1B_2}. \nonumber
\end{eqnarray}
It is immediate that each of the Bell measurements occurs with $25\%$ probability and 
towards $A$ and $C$ the following classical communication protocol is applied
\begin{eqnarray}
 \{\hat{\mathcal P}_1, \hat{\mathcal P}_2\} &\rightarrow& -\ket{\Psi^-}_{AC}, \nonumber \\
 \{\hat{\mathcal P}_1, \hat{\mathcal{I}}-\hat{\mathcal P}_2\} &\rightarrow& e^{i \pi/2} \ket{\Phi^-}_{AC}, \nonumber \\
 \{\hat{\mathcal{I}}-\hat{\mathcal P}_1, \hat{\mathcal P}_2\} &\rightarrow& e^{-i \pi/2}\ket{\Phi^+}_{AC}, \nonumber \\
 \{\hat{\mathcal{I}}-\hat{\mathcal P}_1, \hat{\mathcal{I}}-\hat{\mathcal P}_2\} &\rightarrow& \ket{\Psi^+}_{AC}. \nonumber
\end{eqnarray}

\section{Performance and Limitations}
\label{experiment}

In the subsequent discussion we compare the prerequisite of our proposal with the status of current developments in experimental
physics. In view of these experimental setups and their parameters, we give the performance of our proposed quantum repeater.

\subsection{Experimental considerations}

In our proposal, each repeater node requires four cavities as explained in Sec. \ref{Prot+Setup}, where two of the cavities are 
coupled one sided to fibers. In order that the cavities couple efficiently to the fiber links, they may be built with 
asymmetric mirror transmissions. If mirrors with high transmission rates are coupled to the single-mode optical fiber, then this leads to a highly
directional single-mode output \cite{Ritter}. In this configuration together with$~^{87}$Rb atoms, the whole setup operates in the 
intermediate-coupling regime of cavity QED $\{g, \kappa, \Gamma\} \approx 2\pi \times \{5,3,3\}$ MHz. The other requirement is that these
cavities are also coupled to outer lasers which prepare the coherent states inside them and to balanced homodyne measurement setups. 
A possible solution is to pierce a small hole in the center of the mirror with low transmission rate, thus allowing a good in and out coupling.
This experimental technique has already been reported for microwave cavities \cite{Raimond}. The other two cavities, which are used in the 
entanglement purification protocol and in the generation of complete Bell measurements, have to be able to couple strongly and symmetrically
the supported single-mode radiation field to two atoms. This implementation of the two-atom Tavis-Cummings model has been experimentally
reported for neutral Cs atoms with $\{g, \kappa, \Gamma\} \approx 2\pi \times \{18.0,0.4,5.2\}$ MHz \cite{Reimann}, for$~^{40}$Ca$^+$ ions
with $\{g, \kappa, \Gamma\} \approx 2\pi \times \{1.0,0.05,11.5\}$ MHz \cite{Casabone}, and for$~^{87}$Rb atoms with
$\{g, \kappa, \Gamma\} \approx 2\pi \times \{7.6,2.8,3.0\}$ MHz \cite{Neuzner}. Furthermore, these cavities have to have a good in and out 
coupling with external radiation fields, such that the preparation of coherent state and the postselective field measurements via balanced homodyne
photodetection can effectively be carried out. 

Another critical step in our proposal is that the atoms, implementing the qubits, can be transported through the cavities. The transport 
has to be highly controllable and a possible solution is the use of optical conveyor belts. Cs atoms can be captured from a vapor, cooled down, 
and trapped in a high-gradient magneto-optical trap \cite{Zu}. A dipole trap is formed with help of two counterpropagating laser beams and 
the atoms are transferred without loss from the magneto-optical trap into the dipole trap \cite{Khudaverdyan}. 
By detuning the frequencies of the laser beams one can set into motion the standing wave which acts as a optical conveyor belt and transports 
the atoms with high position precision into the cavities. Furthermore, the speed of the conveyor belts defines the interaction time between
the atoms and the radiation field inside the cavities. Because of the purification protocol, we also require that the number of atoms loaded
in the optical conveyor belt is high as possible ($2^N$ with $N$ iterations) and a decade ago it was reported that it is possible to 
load $19$ atoms efficiently into the dipole trap \cite{Forster}. The atoms in the dipole trap can be subject to coherent manipulations as 
imposed in the purification protocol, in which measurement-dependent transformations have to be carried out (see Sec. \ref{pur}). After the 
qubit-field interactions, information on the internal atomic state have to be extracted. This can be done by applying push-out lasers on 
the conveyor belt and depending on the internal 
atomic state the Cs atom either remain in the trap or get pushed out \cite{Kuhr}. In experiments without conveyor belts  
one may use state-selective field ionization detectors for rubidium atoms \cite{Haroche}. 

Postselective field measurements are at the core of our proposed quantum repeater. These measurements are carried out via 
balanced homodyne photodetection. Here, the purpose is not to perform a complete state tomography on the radiation field  emerging from
the qubit-field interactions, but instead to measure a specific field state component and by thus conditionally postselect qubit states.
All three protocols presented in Sec. \ref{Prot+Setup} depend on the realization of such a measurement, which is capable of discriminating
a coherent state from the rest of the field states, where all states are well separated from each other in the phase space. There are already
investigations for such situations. For example, Ref. \cite{Wittman} has shown that the error probability of a scheme, where $\ket{-\alpha}$ and 
$\ket{\alpha}$ are to be discriminated, is small also for the small mean number of photons involved, i.e., $\bar n \approx 0.4-1.4$. We have considered 
in our scheme $\bar n \approx 100$, so we believe that the implementation of our proposal with postselective field measurements 
is in the range of current experimental technologies and furthermore due to the large number of mean photons involved detector inefficiencies
can also be overcome. The duration time of performing a quadrature measurement depends on the setup. Here, we estimate it to be equal to
$5.5$ ns \cite{Wittman, Hansen, Cooper}.

Optical fibers are the key elements defining the distance between the nodes of a quantum repeater. The critical parameter defining this
distance is the attenuation length, which is maximal at telecom wavelengths around $1.5 \mu$m \cite{Inagaki}. The only issue here is that most of 
the atomic transitions of typical cavity QED atoms, which couple resonantly to the single-mode radiation field, are not at telecom wavelength and 
therefore the field state is not suitable for long-distance transmission over optical fibers due to high losses. 
There are two possible approaches: realizing cascade transitions \cite{Chaneliere, Uphoff} or using wavelength conversion \cite{Eschner}. These
experiments are subject to the generation or conversion of single photons. We remind the reader that our scheme is based on multiphoton field states
and therefore further experimental developments are required by our proposal in order to use optical fibers with telecom wavelength. Otherwise,
resonant photons with atomic transitions will suffer high attenuation in optical fibers with frequencies equal to the atomic transitions. 
Provided that we are able to use optical fibers with telecom wavelength, then the photon loss in our theoretical proposal can be considered to 
be approximately $0.2$ dB/km.  Thus, the propagation time $T$ and the damping rate $\gamma$ can be translated into a length $L_0$,
which characterizes the distance between two repeater nodes, by the relation $L_0=20 (\gamma T)/(0.2 \log 10)$.

\subsection{Rate analysis}

In this subsection we compute the rates at which near-maximally-entangled pairs are generated between the end points of the repeater chain. An 
important parameter is the time $T_{\text{link}}$ which is required to purify a near-maximally-entangled state between two neighboring repeater 
nodes and the overall success probability $P$ of this process.  We are going to estimate these parameters depending on the number of 
elementary links $n$, the number of iterations $N$ required in purification protocol, and the 
success probabilities obtained in Sec. \ref{Prot+Setup}. 

First, we analyze the repeaterless entanglement generation between two points separated by a distance $L_0$. The total time $T_1$ 
attempting to generate an entangled qubit between these two points has the following parts: the time required for the two qubit-field interactions 
$2 \times 1/(2g)$ (characteristic time of the collapse phenomena in the Jaynes-Cummings model); 
the leak-in and leak-out processes $2 \times 1/\kappa$; the propagation time $L_0/c$ where $c \approx 2 \times 10^8$ m/s is the 
speed of light in a telecom optical fiber; the time required for the balanced homodyne measurement $T_{\text{det}}$; and the time
$L_0/c$ of classical communication in order to confirm or deny the success of the procedure. We consider a case, where the 
reinitialization of the cavities and the detectors is done during 
the classical communication. Thus, we obtain
\begin{equation}
T_1=\frac{1}{g}+\frac{2}{\kappa}+\frac{L_0}{c}+T_{\text{det}}+\frac{L_0}{c}.
\end{equation}
Provided that at least two entangled qubit pairs are generated, the purification protocol may start and
the total time $T_2$ attempting to purify a state out of these two states has the following parts: 
time required for the two qubit-field interactions $2 \times 1/(2g)$; the time $T_{\text{det}}$ required for the balanced homodyne measurements and 
the qubit measurements; and the time of classical communication. Here, the classical communication is required both to confirm the 
success of the protocol and to postprocess the obtained state depending on the results of the qubit measurements. These considerations yield
\begin{equation}
T_2=\frac{1}{g}+T_{\text{det}}+\frac{L_0}{c}.
\end{equation}
Now, taking into account the experimental parameters discussed in the previous section, we realize that for distances by means of 
$L_0> 2$ km both $T_1$ and $T_2$ have $L_0/c$ as the dominant time, because $T_{\text{det}}$ is mainly determined by the leaking out of 
the fields from the cavity into the measurement setups, whereas the quadrature measurements are very fast. 
We consider that qubit measurements are also fast. These considerations yield that $T_1 \approx 2L_0/c$ and $T_2 \approx L_0/c$.

If the distance $L_0$ is below $2$ km then $1/\kappa$ is the dominant term in the duration of the process. 
In this case, the reinitialization of the cavities and the leaking of the fields towards the detectors must be taken into consideration 
and we estimate the whole time to be at 
least $2/\kappa$.  Thus, for small distances both $T_1$ and $T_2$ are considered to be approximately equal to $10$ $\mu$s 
($\kappa=2 \pi \times 0.05$ MHz from Ref. \cite{Casabone}).
 
Thus, for a distance $L_0$ (later an elementary link in the repeater chain), the time required to generate $2^N$ entangled 
pairs and obtain a highly entangled pair by $N$ purification rounds is given by 
\begin{equation}
T_{\text{link}}= 2^N T_1+ (2^N-1)T_2, 
\end{equation}
where $2^N-1$ is the number of the purification protocols applied. During this time the overall success probability is
\begin{equation}
 P=\Big[P_{Gen}\Big]^{2^N} P_{Pur}. \label{rateprob}
\end{equation}

\begin{figure*}[t!]
 \begin{center}
 \includegraphics[width=.49\textwidth]{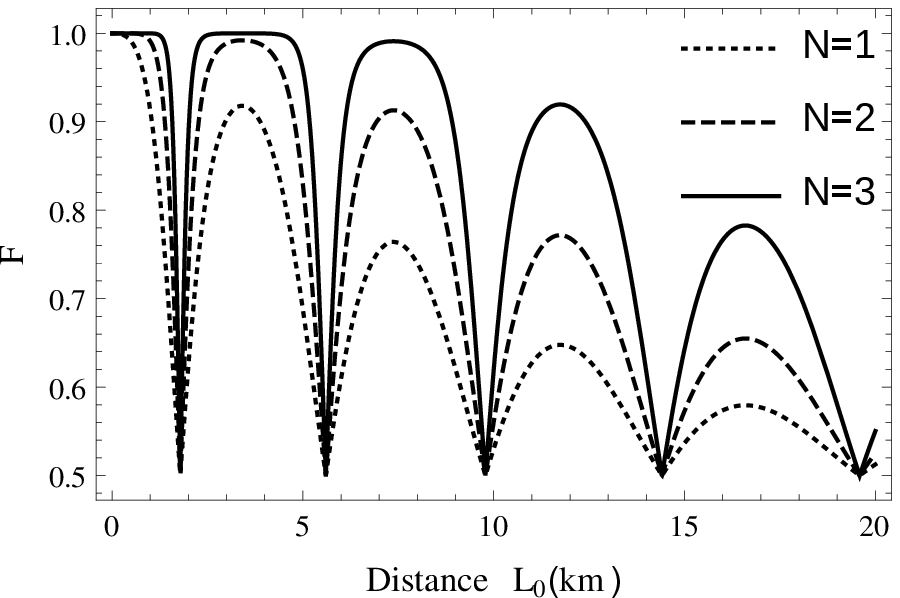}
 \includegraphics[width=.49\textwidth]{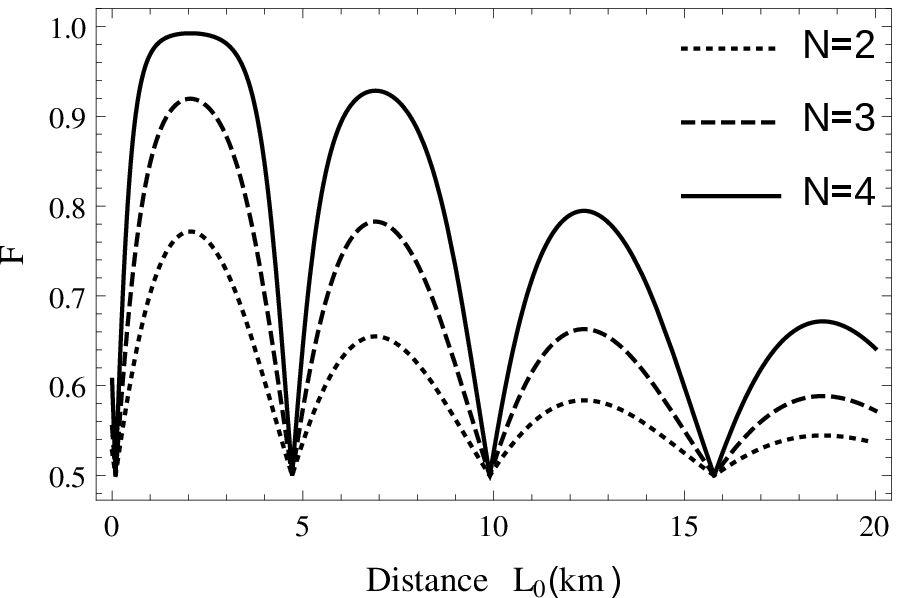}
 \includegraphics[width=.49\textwidth]{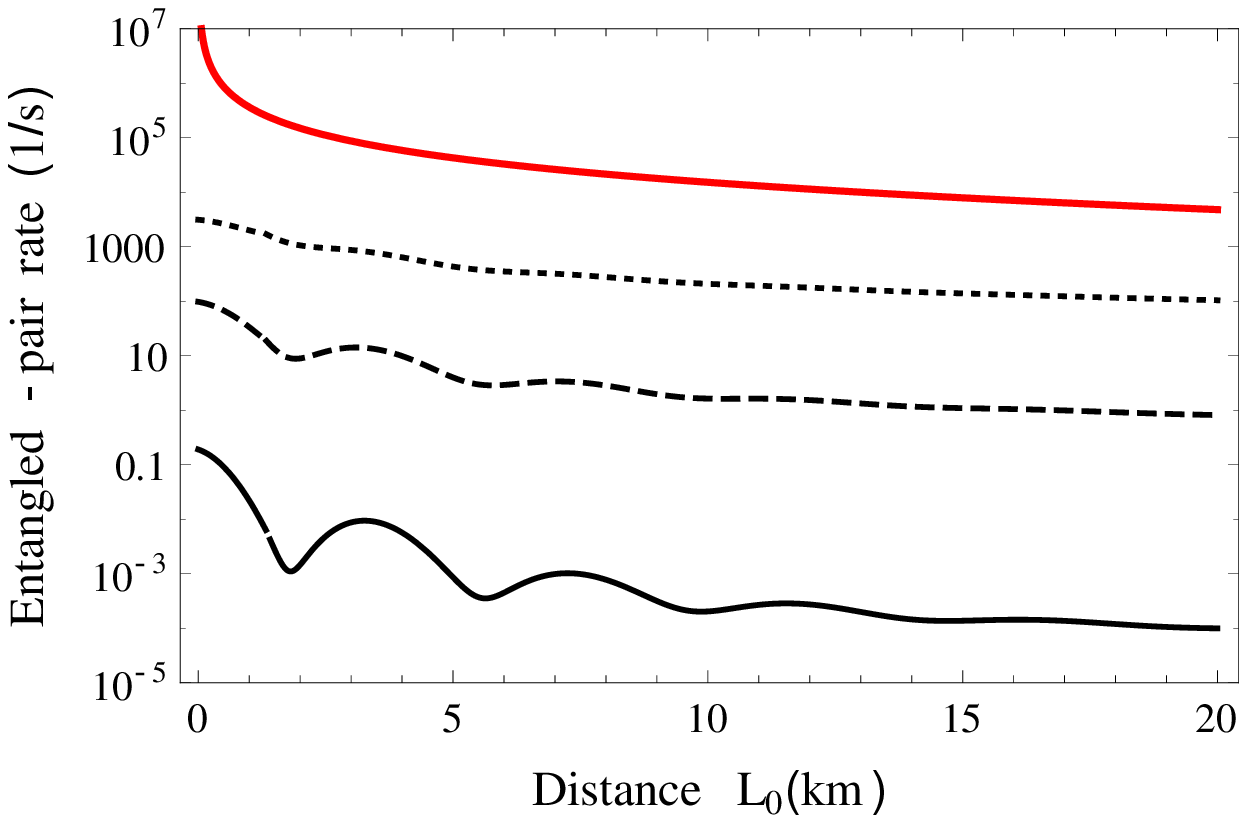}
 \includegraphics[width=.49\textwidth]{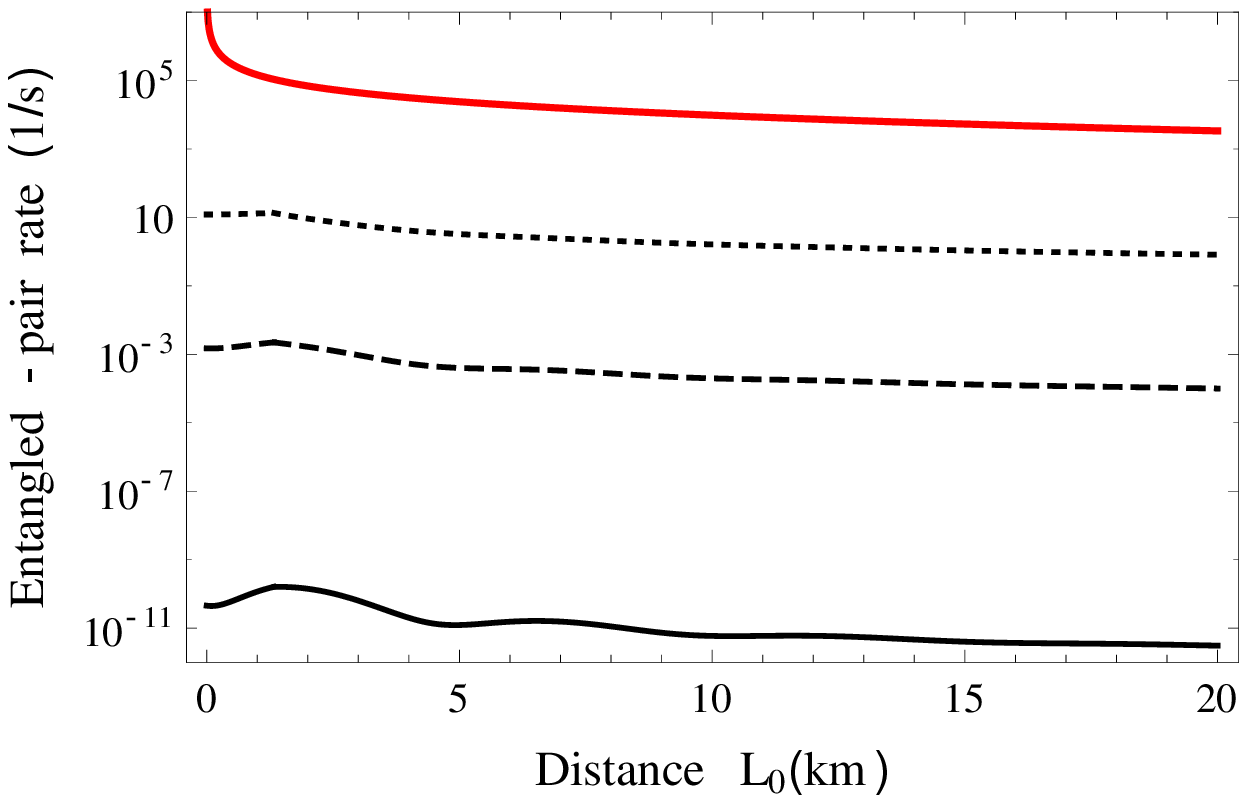}
\caption{Entanglement generation in repeaterless scenarios.
Top panels: Achievable fidelities $F$ (see \eqref{fidforrate}) of entangled pairs generated between two points separated by the
distance $L_0$. $N$ is the number of purification rounds. 
Bottom panels: Semilogarithmic plot of average rates of entangled pairs with the same purification rounds as in the 
same (right or left) top figure. Left panels: The rate $1-\eta$ of reflected number of photons from the surface of the second cavity 
(see Sec. \ref{ent}) is taken 
to be zero. Right panels: $1-\eta=0.2$. The red(gray) curves are the ultimate rates of Bell pairs per second in repeaterless quantum 
communication \cite{Pirandola}, where the transmissivity of the channel with length $L_0$ is defined by pure fiber loss in the left panel and 
fiber loss and $\eta$ in the right panel. Note that not all rate values are accompanied by 
high fidelities in the respective top figure and for isolated cases when $F=0.5$ the pairs are not even entangled.
The interaction time has been set to $\tau=4/g$ with mean photon number $\overline n =100$. }
   \label{fig3}
 \end{center}
\end{figure*}

According to the purification protocol in Sec. \ref{pur}, we purify towards $\ket{\Psi^-}$ or $\ket{\Psi^+}$ depending on the sign of $x$
in Eq. \eqref{entgen}. Therefore, we define the fidelity
\begin{equation}
 F=\max \{f_{(N)},g_{(N)}\} \label{fidforrate}
\end{equation}
where $f_{(N)}$ and $g_{(N)}$ are given in Eq. \eqref{purifyN}, i.e., the general form of a purified state after $N$ purification rounds. 
Applying our protocols of 
entanglement generation and purification to a point-to-point situation, one may talk about the rate of entangled pairs generated across a 
distance $L_0$ only if the achievable fidelities are also displayed as well. In this case, we are able to compare the performance of our scheme
to a recent results of Ref. \cite{Pirandola}, where the ultimate rate of repeaterless quantum communication is given by $-\log_2(1-\chi)$. 
This formula gives the rate of Bell pairs per channel use, which results in the rate of Bell pairs per second by multiplying it
with the repetition rate $c/(2L_0)$; i.e., entanglement generation over the quantum channel is assisted by a classical communication.
The parameter $\chi$ quantifies the fraction of photons surviving the channel, i.e., the transmissivity of the channel, 
which in our case yields $\chi=\eta \exp\{-\gamma T\}$. $\eta$
and $\gamma$ have been introduced in Sec. \ref{ent} to characterize the fraction of photons not reflected
from the surface of the second cavity and the decay rate of the optical fiber. 

In Fig. \ref{fig3}, average rates of generated
entangled pairs over a distance $L_0$ are analyzed for different numbers $N \in \{1,2,3,4\}$ of purification rounds and   
$\eta \in \{0.8, 1\}$. These figures clearly demonstrate that there is a tradeoff between average rates and the fidelity of the pairs with respect
to a Bell state ($\ket{\Psi^-}$ or $\ket{\Psi^+}$). If $\eta=1$, then near-maximally-entangled pairs are either generated on 
very short distances $L_0\approx 500$ m with an average rate $\approx 2625$ pairs per second or larger distances $L_0\approx 4$ km with
an average rate $\approx 10^{-2}$ pairs per second. If $\eta=0.8$, then we require at least $N=4$ purification rounds and for a distance of
$L_0\approx 2.5$ km we obtain a very low average rate $\approx 10^{-11}$ pairs per second. These distances and average rates define
also the possible elementary links of the repeater chain, because applying entanglement swapping procedure to low-fidelity pairs reduces
the fidelity of the output pairs even further. Now, if we compare our results with the ultimate rate of Ref. \cite{Pirandola}, it becomes clear
that our protocol has a low performance unless $\eta \approx 1$ and $L_0\leqslant0.5$ km. However, the result in Ref. \cite{Pirandola} is an 
upper bound for rates assuming arbitrary local operations and unlimited classical communication, thus being a benchmark rate for quantum 
repeater proposals.

In the next step, we are going to discuss several scenarios where the quantum repeater protocol is in use. Let us denote by $L=nL_0$ 
the length of the repeater chain with $L_0$ being the length of an elementary link and consequently $n$ being the number of the links.
We calculate the average number of 
attempts of preparing one near-maximally-entangled pair between all the repeater nodes.
The best strategy here is to use memories and implement $n$ parallel processes. As soon as one near-maximally-entangled pair
has been generated along one elementary link, its state is saved in a quantum memory, while between the other nodes the process is 
repeated until we succeed. Provided that we are successful along all the links, the average number of attempt is \cite{Bernardes}
\begin{equation}
 A_n=\sum^n_{i=1} \binom ni \frac{(-1)^{i+1}}{1-(1-P)^i} \label{An}
\end{equation}
with $P$ given in Eq. \eqref{rateprob}.

Another characteristic time of a quantum repeater, which is $T_{\text{swap}}$, stands for the overall time required to 
entangle the two end points of the repeater chain with the help of entanglement swapping procedures. Based on our proposal in 
Sec. \ref{swap}, the swapping procedure is 
deterministic and the characteristic time of a single swap contains the following processes: the time required for the 
two subsequent qubit-field interactions $2 \times 1/(\sqrt{2}g)$ (characteristic time of the collapse phenomena in the two-qubit 
Tavis-Cummings model); the time $T_{\text{det}}$ required for the two balanced homodyne measurements; 
and the time of classical communication between the nodes in order to inform the parties about which state has been swapped. In order to 
speed up the whole swapping process, parallel entanglement swappings are carried out for intermediate nodes. For example, $n=100$ elementary
links are reduced in the first round to $n=50$ links, in the second round to $n=26$ links, and so on, until we have $n=1$ link, 
which means that we reached the end points of the repeater chain. Hence,
\begin{equation}
 T_{\text{swap}}= \lceil \log_2 n \rceil \left(\frac{\sqrt{2}}{g}+ 2 T_{\text{det}} + \frac{L_0}{c} \right),
\end{equation}
where $\lceil . \rceil$ is the ceiling function. The time $T_{\text{swap}}$ has the dominant term $L_0/c$ unless $L_0 < 2$ km, when
we consider $T_{\text{swap}}=10$ $\mu$s, where $g=2 \pi \times 1.0$ MHz and $T_{\text{det}} \approx 1/\kappa$ with $\kappa=2 \pi \times 0.05$ MHz 
(see Ref. \cite{Casabone}).  

An interesting feature arises when we apply the swapping procedure to $\hat{\rho}_{A,B_1} \otimes \hat{\rho}_{B_2,C}$ with Bell measurements
on qubit systems $B_1$ and $B_2$ and the shared state between the nodes being the 
output state of the purification protocol (see Eq. \eqref{purifyN})
\begin{equation}
\hat{\rho}=p \ket{\Psi^-}\bra{\Psi^-} +  (1-p) \ket{\Psi^+}\bra{\Psi^+}. \nonumber
\end{equation}
We have iterated the entanglement purification until either 
$p$ or $1-p$ is larger than $0.999$; however, the equality $p=1$ or $p=0$ in principle can not be reached in finite numbers of 
purification rounds. During the applications of the swapping procedure, this is an issue, because the fidelity $F$ introduced in 
Eq. \eqref{fidforrate} reduces after $k$ rounds of parallel swapping procedures as
\begin{eqnarray}
 F^{(k)}&=&f\big(F^{(k-1)}\big), \quad F^{(0)}=F, \label{swapred} \\
 f(x)&=&1-2x+2x^2. \nonumber
\end{eqnarray}
As an example, consider a repeater chain with $n=60$ elementary links, which means that the number of parallel swapping procedures
$k=\lceil \log_2 60 \rceil=6$. If the fidelity of the pairs between the nodes is $F=0.999$, then the fidelity of the final pair between the end points
is $F^{(6)}=0.939$. Therefore, in order that we can talk about near-maximally-entangled pairs ($F>0.999$), two more purification rounds 
have to be performed at the end 
points of the repeater chain. In numbers, an average of $41$ pairs have to be postprocessed. Therefore, we introduce the average number of pairs  
$\bar N=2^j/ P_{\mathrm{Pur}}$ (see Eq. \eqref{purprob} for $j$ purification rounds) involved in the final purification procedure, where $j$ 
depends on $F^{(k)}$ (see Eq. \eqref{swapred}) with
$k=\lceil \log_2 n \rceil$ and $n$ is the number of elementary links.

Finally, the average rate is given by
\begin{equation}
 R=\frac{1}{\bar N \left(T_{\text{link}}A_n + T_{\text{swap}}\right)}. \label{rate}
\end{equation}

\begin{figure}[t!]
 \begin{center}
 \includegraphics[width=.5\textwidth]{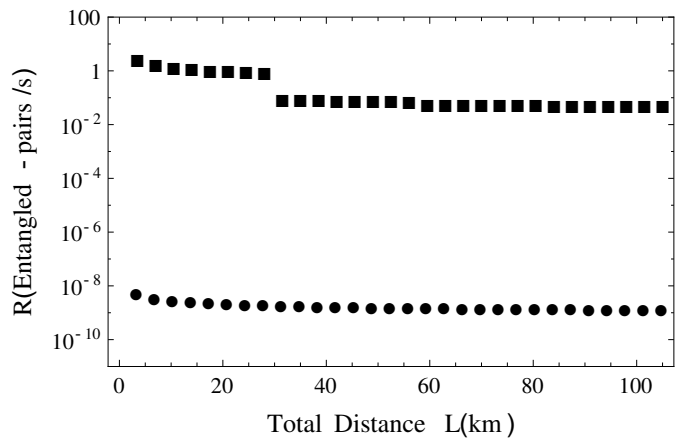}
\caption{
Semilogarithmic plot of average rates of near-maximally-entangled ($F>0.999$) pairs between two end points separated by the total
distance $L$. The elementary link between 
two nodes is $3.5$ km. Two scenarios are presented for 
different number of iterations of the purification protocol: $N=2$ (points displayed as squares) and $N=4$ (circles). Drops in the average rate
are due to the destructive effects of the swapping procedure on the fidelities, 
according to Eq. \eqref{swapred}, and therefore extra entanglement purification is required at the end points of the repeater chain. 
$1-\eta=0$; i.e., there are no photons reflected back from the surface of the second cavity (see Sec. \ref{ent}), and 
the elementary link length determines accordingly the fidelity of the repeater protocol's input pairs in Eq. \eqref{entgen}.}
   \label{fig4}
 \end{center}
\end{figure}

We have already mentioned that the number of qubits available at the nodes is limited due to the current stage technology of
conveyor belts. As this number is $19$ (see Ref. \cite{Forster}) we will consider no more than four iterations of the entanglement 
purification procedure. Although this number is small, Fig. \ref{fig4} shows that four iterations decrease extremely 
the average rate $R$ of generated near-maximally-entangled pairs. This results clearly reflects the very expensive nature of purification protocols and 
shows that increasing
the number of purification rounds leads to unrealistic demands of quantum memory. In the case of $N=4$, this means that
we require a quantum memory which is capable of protecting the coherency of the states for $10^{8} s$, i.e., more than $3$ years. In the $N=2$ case,
we have much higher average rates, however the fidelity of the states (still larger than $0.999$) obtained after two purification rounds 
is affected by the swapping procedures and a few more 
purification rounds have to be carried out at the end points of the repeater chain. 
Here, we have considered an ideal scenario where  $\eta = 1$ (see Eq. \eqref{eta}) and $x=-0.5$, which according to 
Fig. \ref{fig1} yields an elementary link length $L_0 \approx 3.5$ km.

As we increase the number of elementary links in order to 
obtain larger distances $L$, we are facing a situation where the fidelities of the pairs are more reduced by the swapping procedures. 
Thus, extra
purification rounds are required at the end points of the repeater chain and this post-process results in the decrease of the average rates. 
For example, let us consider that across all links we have purified pairs with fidelity $F=1-\epsilon$, where 
$\epsilon$ is a threshold number defining what we call a near-maximally-entangled pair 
($\epsilon<0.001$ in this paper). If we have $n$ elementary links then the fidelity after the swapping procedures is approximately equal to 
$1-2^{\lceil \log_2 n \rceil} \epsilon$. Therefore, strategies for large distances and with lower number of purification rounds may not 
generate pairs with sufficiently high fidelity, such that these fidelities do not drop way below the threshold fidelity $1-\epsilon$ after the 
swapping procedures are applied. Depending how much they have dropped, more purification rounds have to be carried out at the end points of
the repeater chain, which yield significantly reduced average rates. These reduced rates may be comparable with other strategies 
with high number of purification rounds.

\begin{figure}[t!]
 \begin{center}
 \includegraphics[width=.5\textwidth]{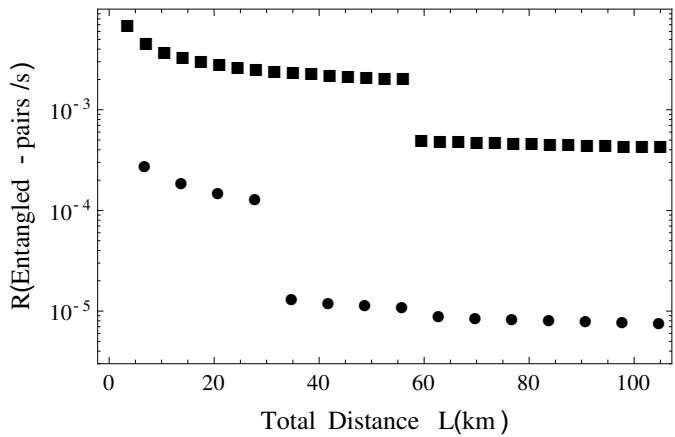}
\caption{
Semilogarithmic plot of average rates of near-maximally-entangled ($F>0.999$) pairs between two end points separated by the total
distance $L$. Two cases are presented for 
different elementary link lengths: $L_0=3.5$ km (points displayed as squares), and $L_0=7$ km (circles). Both cases are depicted for 
$N=3$ purification rounds. For a brief explanation for the drops in the average rates, see Fig. \ref{fig4}. 
$1-\eta=0$; i.e., there are no photons reflected back from the surface of the second cavity (see Sec. \ref{ent}), and 
the elementary link lengths determine accordingly the fidelities of the repeater protocol's input pairs in Eq. \eqref{entgen}.}
   \label{fig5}
 \end{center}
\end{figure}

In Fig. \ref{fig5}, we compare two quantum repeaters with the only difference being the elementary link length.  We have considered a scenario
where $\eta=1$ and we have taken for $x$: $-0.5$, i.e., $L_0 \approx 3.5$ km, and $0.3$ , i.e., $L_0 \approx 7$ km
(see Fig. \ref{fig1}). Despite the longer distance of the elementary link length, low probabilities of purifying the state characterized 
by $x=0.3$ reduce the average rate $R$ of generated near-maximally-entangled pairs at the two end points separated by the total distance $L$. Furthermore, the destructive
effects of the swapping procedures affect both strategies. The scenario with shorter elementary links is less affected, because here 
we generate higher fidelity pairs between the repeater nodes than in the scenario with $7$-km-long elementary links.

\begin{figure}[t!]
 \begin{center}
 \includegraphics[width=.5\textwidth]{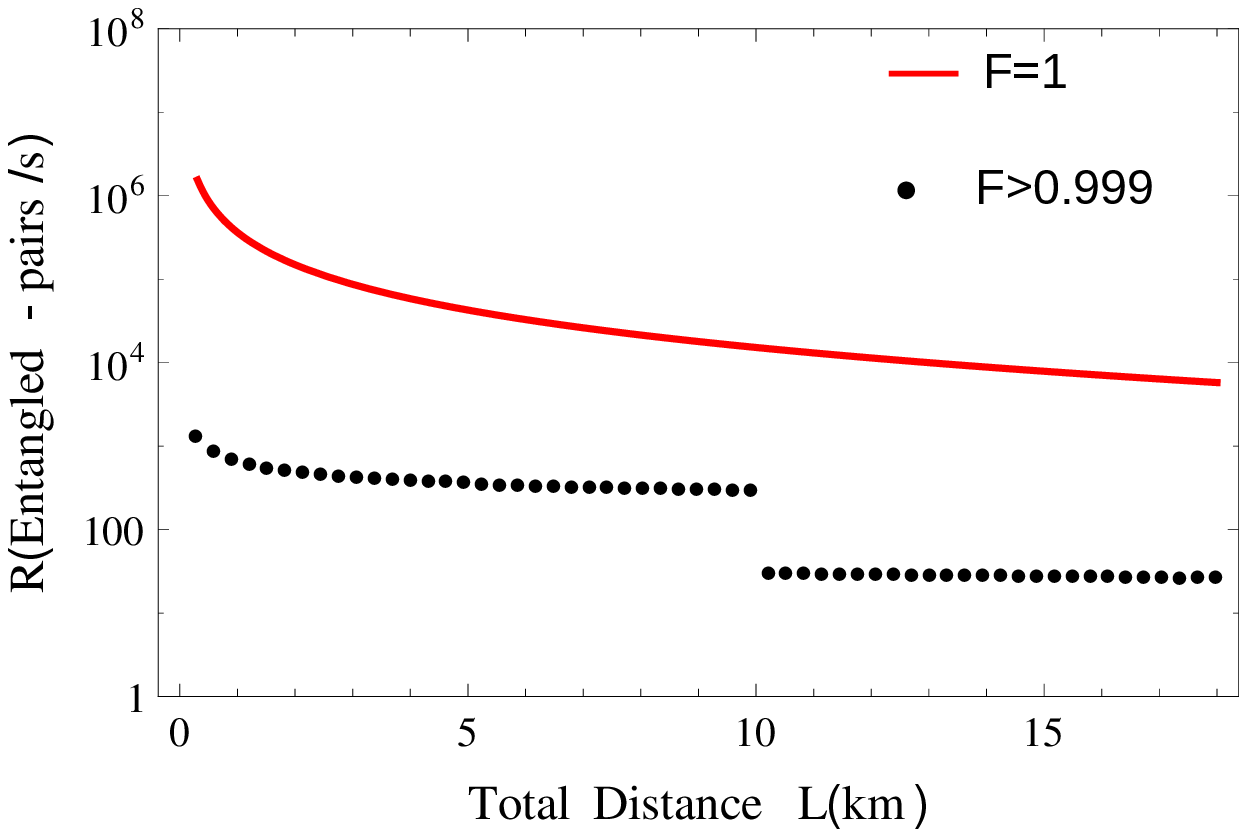}
 \includegraphics[width=.5\textwidth]{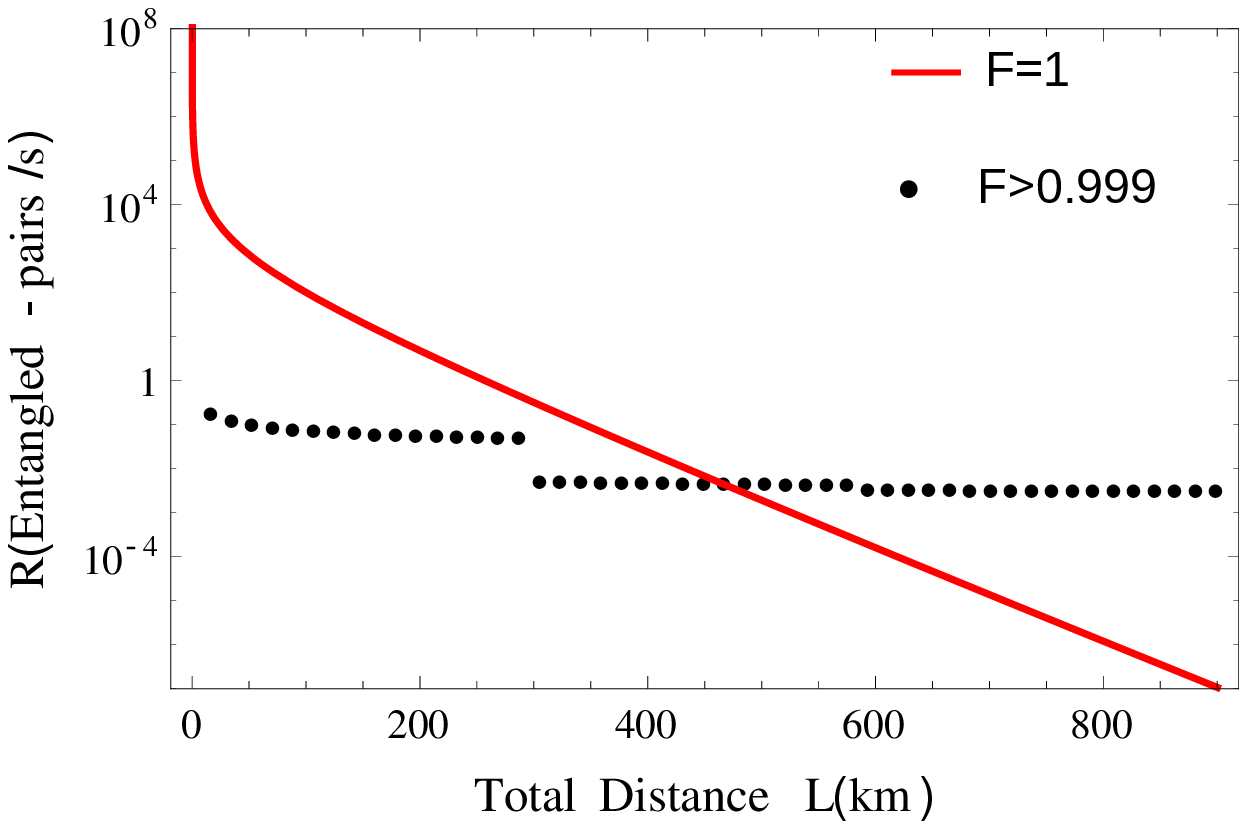}
\caption{
Semilogarithmic plot of average rates of entangled pairs between two end points separated by the total
distance $L$. Top panel: The scenario (black dots) with $L_0=0.3$ km long elementary links and $N=1$ purification round 
is depicted up to $18$ km, where near-maximally-entangled ($F>0.999$) pairs are generated. Bottom panel: The $18$-km-long repeater
chain of the top figure is considered now as an elementary link and with $N=1$ purification round is depicted up to $900$ km. 
$1-\eta=0$; i.e., there are no photons reflected back from the surface of the second cavity (see Sec. \ref{ent}), and 
the elementary link length determines accordingly the fidelity of the repeater protocol's input pairs in Eq. \eqref{entgen}. 
The solid lines with $\eta=1$ are the ultimate rates of pure fiber loss in repeaterless quantum communication \cite{Pirandola}.  
}
   \label{fig6}
 \end{center}
\end{figure}

In the last case, we set $\eta=1$ and $L_0=0.3$ km, which yield $x=0.913$. This means that the fidelity of the generated pairs
(see Eq. \eqref{entgen}) is high enough to obtain a near-maximally-entangled pair after only one round of the purification protocol. 
In the top panel of Fig. \ref{fig6}, we compare this scenario with the ultimate rate of Bell pairs per second in repeaterless quantum communication \cite{Pirandola} 
and we see that the average rate of the quantum repeater protocol starts with lower values, but it seems that scales with $L$ better.
In this scenario, we are bound in our numerics to $18$ km, because we have $60$ elementary links, and the binomial
$\binom ni$ in \eqref{An} may take extremely large values which are multiplied with very small numbers. Above $18$ km or $60$ elementary links 
the numerical instabilities are increasing and they lead to meaningless average rate numbers. It is also worth mentioning that in this 
case with $60$ elementary links one must build $59$ intermediate repeater nodes, which is an expensive procedure in 
regard to physical resources of cavity QED. In order to show that this repeater scenario exceeds the ultimate limit of Ref. \cite{Pirandola},
we embed the above discussed of an $18$-km-long repeater chain into a longer repeater chain as an elementary link. Therefore, 
we determine the probability of generating an entangled pair with $x=0.8769$ over $18$km, which yields $0.026$. This approach circumvents
the use of large valued binomials $\binom ni$. In the bottom panel
of Fig. \ref{fig6}, we see that this longer repeater chain crosses the ultimate repeaterless rate around $500$ km and for a total distance of
$L=900$km the rate is found to be $3.6 \times 10^{-3}$. This achievement is contrasted with the required number of repeater nodes, which turns out
to be $2999$. 

We therefore conclude that for a realistic implementation of our proposal, the generation of high-fidelity entangled states between the nodes 
is the most crucial ingredient. Entanglement purification is a very expensive procedure which can not be properly compensated for by high repetition
rates and the best strategy would be to generate entangled states which can be purified in one step \cite{TB}. Our numerical 
investigations shows that the parameter $\eta$ (defined in Eq. \eqref{eta}) quantifying the fraction of photons entering from the optical 
fiber into the cavity is the most crucial hurdle for obtaining high average rates, because the photon loss rate $\gamma$ can be 
recompensated by choosing shorter elementary links. There exists
experiments, which are able to obtain $\eta \approx 1$; however, they operate with single photons on short distances \cite{Ritter}. In scenarios,
where $\eta=1$, we obtained an average rate $R=23$ with $60$ elementary links for $L=18$ km and $R=3.6 \times 10^{-3}$ with $3000$ elementary 
links for $L=900$km. Furthermore, this scenario from a theoretical point of view outperforms the ultimate rate of repeaterless quantum 
communication \cite{Pirandola}. A much lesser average rate $R=5\times10^{-4}$ 
is found for $30 $ elementary links and a  total distance $L=105$ km. In summary, large distances and 
high repetition rates require a lot of repeater nodes at the expense of physical resources. 

\section{Conclusions}
\label{conclusions}

We have presented a hybrid quantum repeater based on resonant qubit-field interactions. In our scheme, all 
two-qubit operations required for the building blocks are generated via qubit-field interactions and postselective field measurements, thus
making our proposal a good candidate for experimental implementation. 

In the context of entanglement generation between the repeater nodes, we have investigated a system of two spatially separated
material qubits coupled to single-mode cavities. In addition, these cavities are connected by an optical fiber. 
For the description of qubit-field interactions, we have used the resonant Jaynes-Cummings-Paul model and entanglement is generated between 
the distant qubits by a postselective balanced homodyne photodetection. Our model is subject to two type of decoherence, 
namely the photon loss in the optical fiber and the photon reflection from the surface of the cavity, an effect of the fiber-cavity 
coupling inefficiencies. These considerations extend former studies on hybrid quantum repeaters. Within this model, we have found that the 
quality of entangled qubit states, quantified via the concurrence, is very sensitive to the the photon reflection, 
which has a strong impact on the orthogonality of field states 
involved in postselective field measurements. In the case of small reflectivity and several-km-long optical fibers, we have shown that
high-fidelity entangled states can be created with $50\%$ success probability, which is an improvement of our former 
result in Ref. \cite{Bernad1}. 

In the next step, entangled state obtained in the first building block of the quantum repeater have 
been considered as input states for an entanglement purification protocol, a recurrence protocol, introduced by us \cite{Ludwig, TB}. The 
theoretical model consists of two qubits, which sequentially interact with a single-mode cavity, and postselective field measurements. We have used
the Jaynes-Cummings-Paul model and its solutions for the description of the interactions. These interactions and the field measurements generate a 
probabilistic two-qubit quantum operation, which takes over the role of the controlled-NOT gate used in standard purification protocols. 
We have found that the overall success probability of purifying near-maximally-entangled pairs is very low, and the results for four 
steps of iterations are already unrealistic. Therefore, it is more beneficial to use as few 
purification rounds as possible; otherwise the entanglement protocol becomes very expensive in regard to physical resources.  

For the final building block of the quantum repeater, the entanglement swapping, we have considered two qubits, 
which interact simultaneously with single-mode cavities. We have employed the Tavis-Cummings model and its solutions.
The Bell measurement are generated by postselecting the emerged cavity fields. This study, which is based on our former results 
\cite{Torres, TorresA}, has been extended by a new set of field measurements, which are able to project on field states lying on the opposite 
side of phase space's axes. These postselective field measurements allow for deterministic realization of unambiguous Bell measurements, 
an improvement of our result in Ref. \cite{TorresA} and an important ingredient in the effective actuation of quantum repeaters.

As all three building blocks consists of the same cavity QED elements, we have collected some recent experimental developments with respect to 
these components. We have presented the parameters of these experiments and discussed the pros and cons of an implementation.
A comprehensive analysis of the quantum repeater's rate of generating near-maximally-entangled pairs per second
has been given. In particular, we have found that moderately low rates can be achieved in the context of current experimental technologies for 
distances up to $100$ km. This result is mainly due to the request that we purify near-maximally-entangled pairs ($0.999<F<1$) between the neighboring nodes. 
In addition, these purified pairs can not reach in principle $F=1$ and therefore the swapping procedures have destructive effects on the fidelities. 
This may result in extra purification rounds for the entangled pairs between the end points of the repeater chain. 
If the purification protocol is required at the end points, then the average rates are further reduced. 
We have also compared our results with the ultimate rate of repeaterless quantum communication \cite{Pirandola} and we have shown 
that the average rates of our proposal with very high number of nodes exceeds this benchmark 
value around $500$ km. This occurs in cases where the elementary links are a few hundred meters long and we use only one purification round. 

In summary, the strength of our proposal is in the compatible and cavity-QED-based building blocks, which can easily augment each other. 
The main idealistic assumptions throughout this paper are the following: nondecaying qubits, i.e., perfect quantum memories, and 
unit efficiency detectors. In future work, we aim to relax one or both of these assumptions. In view of these considerations, 
our proposal gives a better understanding of the influence of the building blocks on each other and shows its own limitations on 
the achievable repeater rates on moderate distances. These limitations may be surpassed only if in the first building block 
we generate such type of entangled states, which can be purified into a Bell state ($F=1$) in one purification round \cite{TB}. 
Hence, this scenario is able to avoid the low success probabilities of several purification rounds and the 
destructive effects of swapping procedures.

We hope that our work is a step forward to an experimental realization of the first hybrid quantum repeater. In addition, the proposed scheme
mainly relies on current technology and thus offers a clear perspective on a future experimental demonstration.  

\begin{acknowledgments}
This work is supported by the Bundesministerium f\"ur Bildung und Forschung project Q.com and 
the European Union’s Horizon 2020 research and innovation programme under Grant Agreement No. 732894 (FET Proactive HOT). 
In addition, the author would like to thank the anonymous reviewer for helpful and constructive
comments that greatly contributed to improving the paper.
\end{acknowledgments}


\begin{thebibliography}{99}
\bibitem{commentus} Exponentially low probabilities are due to the constraint that at every iteration of the purification protocol the 
two-qubit operations on all the qubit pairs have to be successful. In Fig. $5$ of our former work \cite{Ludwig}, 
the overall success probability was erroneously overestimated and therefore the average number of qubits was underestimated.
\end{thebibliography}
\end{document}